\documentclass[]{mn2e}

\usepackage{graphicx}
\usepackage{txfonts}
   \title[]{Radiation fields in star-forming galaxies: the disk, thin disk and 
bulge\thanks{The data are available in electronic form at the CDS via
  anonymous ftp to cdsarc.u-strasbg.fr}}

   \author[C. C. Popescu and R.J. Tuffs]{Cristina
     C. Popescu$^{1}$\thanks{E-mail: cpopescu@uclan.ac.uk}
          and
          Richard J. Tuffs$^{2}$\thanks{E-mail: Richard.Tuffs@mpi-hd.mpg.de}\\
   $^1$ Jeremiah Horrocks Institute, 
              University of Central Lancashire,
              PR1 2HE, Preston, UK\\
   $^2$ Max Planck Institut f\"ur Kernphysik, Saupfercheckweg 1, D-69117
             Heidelberg, Germany}
\begin{document}

\date{Accepted 2013 September 1. Received 2013 September 1; in original form
  2013 March 25}

\pagerange{\pageref{firstpage}--\pageref{lastpage}} \pubyear{}

\maketitle

\label{firstpage}
\begin{abstract} 
     We provide and describe a library of diffuse stellar radiation fields in 
     spiral galaxies derived using calculations of the transfer of stellar 
     radiation from the main morphological components - disks, thin disks, and 
     bulges - through the dusty interstellar medium. These radiation 
     fields are self-consistent with the solutions for the integrated 
     panchromatic spectral energy distributions (SEDs) previously presented 
     using the same model. Because of this, observables  calculated from the 
     radiation fields, such as gamma-ray or radio emission, can be 
     self-consistently combined with the solutions for the 
     ultraviolet/optical/submillimeter SEDs, thus expanding the range of 
     applicability of the radiation transfer model to a broader range of 
     wavelengths and physical quantities. We also give analytic solutions for 
     radiation fields in optically thin stellar disks, in stellar 
     disks with one dust disk and in stellar disks with two dust disks. The 
     analytic solutions for the direct light are exact and can be used as 
     benchmarks. The analytic solutions with scattering are only approximate, 
     becoming exact only in the extreme optically thick limit. We find 
     strongly contrasting solutions for the spatial distribution of the 
     radiation fields for disks, thin disks and bulges. For bulges we
     find a strong dependence of the radiation fields on S\'{e}rsic index.
\end{abstract}


%

\section{Introduction}

Knowledge of radiation fields (RF) in galaxies is of 
prime importance to many branches of astrophysics, embracing low-
and high-energy processes in the interstellar and intergalactic medium,
with applications to cosmology, galactic astrophysics, astro-chemistry
and astro-biology. For example ultraviolet (UV) radiation fields are one of 
the most important heating agents for interstellar gas, proceeding either 
through photoionisation in the presence of ionising UV radiation, or through the
photoelectric effect in the neutral ISM. Thus, the main cooling
lines of the interstellar medium, in particular the [CII] line, are produced as
a result of the photoelectric effect on dust grains, primary driven by
non-ionising UV photons from the ambient radiation fields interacting with
polycyclic aromatic hydrocarbon (PAH)
molecules and dust grains, both in star-formation regions (e.g. 
Crawford et al. 1985,
Tielens \& Hollenbach 1985, Wolfire et al. 1989,
Hollenbach 1991, Israel \& Maloney 2011) and in the 
diffuse ISM (e.g. Madden et al. 1993, Bennett et al. 1994, 
Pierini et al. 1999, Pierini et al. 2001, Rubin et al. 2009, Lebouteiller et
al. 2012). Diffuse RFs impinging on molecular clouds in the
interstellar medium also drive the excitation of molecules in the outer layers
of the clouds, determining the rotational-line emission
from rare tracers such as CO molecules, from which the 
bulk of the mass in the form of molecular hydrogen can be determined. 
A quantitative knowledge of the
amplitude and spectrum of the RFs is also a necessary input for hydro-dynamical 
simulations of the turbulent structure of the interstellar medium
(e.g. de Avillez \& Breitschwert 2007). 

In high-energy astrophysics it is particularly important to know the number
density of all photons, including those of low energies where the number 
density is high, because such photons can be transformed into gamma-rays
through the inverse Compton scattering (ICS) with relativistic electrons. 
Indeed, inverse Compton scattering provides one of the principal $\gamma$-ray 
production
mechanisms (Jones 1968, Blumenthal \& Gould 1970, Aharonian \&
Atoyan 1981, Narginen \& Putanen 1993, Brunetti 2000, 
Sazonov \& Sunyaev 2000). In a variety of astrophysics environments, from very
compact objects like pulsars and active galactic nuclei to extended sources
like supernova remnants and clusters of galaxies, low-energy photons are
efficiently boosted to high energies through this mechanism (Aharonian \&
Ambartsumyan 1985, Zdziarski et al. 1989, Dermer \& Atoyan 2002, Moderski et
al. 2005, Khangulyan \& Aharonian 2005, Kusunose \& Takahara 2005, Stawarz et
al. 2006, 2010). In galaxies 
the UV/optical and far-infrared (FIR) interstellar radiation fields
can be an important channel for the production of inverse Compton
$\gamma$-rays.  Thus, knowledge of the interstellar RFs is needed to measure the
distribution of relativistic electrons in these objects, as well as to isolate
other gamma-ray emissions, such as those produced in the decay of pions
generated in the interaction of relativistic protons with interstellar gas.
Similarly, because relativistic electrons also produce radio synchrotron
emission, knowledge of the RF in galaxies is necessary to elucidate
the strength of interstellar magnetic fields from a combination of
radio and gamma-ray observations.
Radiation fields also affect the propagation of gamma-rays through the
intergalactic medium (e.g. Stecker et al. 1992) as well as even in the 
interstellar medium
(Moskalenko et al. 2006) through pair production of gamma-ray photons on the 
background radiation fields. 

In summary it is clear that the observational signatures of many of the most 
fundamental physical attributes of
galaxies, including the mass of interstellar gas, in both atomic and molecular
form, the thermodynamical properties of this gas, as well as the ambient
magnetic fields and cosmic rays in the form of both protons and nuclei and
electrons, depend on the radiation fields in galaxies. It would be relatively
straightforward to determine radiation fields by direct observations of the
sources of photons, which, in most cases, are predominantly stars, if 
galaxies were optically thin. However,
galaxies - and in particular star-forming galaxies - contain dust, which,
because it absorbs and scatters photons,
partially or wholly prevents a direct measurement of the spatial and spectral
distribution of the sources of the RFs. The propagation of light depends in a
complex way on the relative distribution of stellar emissivity and dust
opacity, with structures ranging in scale from parsecs to
kiloparsecs. Fortunately, although dust is the cause of this problem, it is also
part of the solution, because the absorbed light is re-radiated in the
mid-infrared (MIR)/FIR/submillimeter (submm). In particular grains which are 
sufficiently large to be in 
equilibrium with the radiation fields directly probe the strength of these radiation fields, acting as a
strong constraint on the radiation transfer problem. In addition, the radiation
emitted by grains small
enough to be impulsively heated by individual photons strongly depends on
the colour of these photons, placing particular constraints on the UV component
of the interstellar radiation fields, providing one has independent knowledge
of the size distributions of the constituent populations of grains. 

Thus, the key to determining radiation fields in galaxies is to invert the
observed broad-band spectral energy distribution incorporating both
measurements of direct light in the ultraviolet (UV)/optical, as well as the
dust re-radiated light in the Far-infrared (FIR)/submillimeter (submm), 
self-consistently taking into account constraints on the spatial distribution of
both stars and dust as well as on the optical properties of the grains. This is the only 
way in which the exact 
contribution of optical and UV photons in heating dust grains can be traced,
allowing for a proper determination of the contribution of old and young
stellar populations in heating the dust (Popescu \& Tuffs 2010), and
therefore allowing for a consistent 
estimation of the star-formation rates in galaxies, and, ultimately
for a robust determination of their 
star-formation history.  
Self-consistent modelling of the panchromatic 
spectral energy distributions (SEDs) of galaxies (e.g. Rowan-Robinson \&
Crawford 1989, Rowan-Robinson 1992, Silva et al. 1998, 
Popescu et al. 2000, Efstathiou, Rowan-Robinson \& Siebenmorgen 2000, 
Misiriotis et al. 2001, Efstathiou \& Rowan-Robinson 2003, Popescu et al. 2004, Siebenmorgen \& Kr\"ugel 2007, 
Rowan-Robinson \& Efstathiou 2009, Bianchi 2008, Baes et al. 2010, Baes et al. 2011, Popescu et al. 2011, 
MacLachlan et al. 2011, Schechtman-Rook et al. 2012, de Looze et al. 2012a,b) 
have been primarily motivated by the need to extract intrinsic parameters of 
the stars and dust, most notably the  aforementioned SFRs, SF histories and 
dust masses (see also review of Rowan-Robinson 2012).  

In previous papers (Popescu et
al. 2000, Tuffs et al. 2004, Popescu et al. 2011) we provided
self-consistent solutions for modelling the integrated panchromatic SEDs of
star-forming disk galaxies. Here we
give and describe the solutions for the radiation fields, 
implicit to the SEDs presented in our previous work. The solutions are given in
the form of a library of stellar radiation fields for the main
morphological components of our model of spiral galaxies: disks, thin disks 
and bulges. These radiation fields are identical to those used when 
calculating the library of panchromatic SEDs presented in
Popescu et al. (2011). Because of this observables calculated from the
radiation fields, such as gamma-ray or radio emission, can be self-consistently
combined with the solutions for the UV/optical/submm SEDs found using Popescu
et al. (2011), thus expanding the range of applicability of the radiation
transfer model to a broader range of wavelengths and physical quantities. 
Accordingly, the library of radiation fields is calculated for
different values of central face-on dust opacity, as defined in Popescu et
al. (2011). The separate tabulation of the radiation fields for disks, 
thin disks and bulges is possible because all the 
stellar components are attenuated by the same distribution of dust, and
therefore the diffuse radiation fields seen by each grain can
be considered to be a sum of the radiation fields produced by each stellar
component. Thus, the radiation fields can be combined according to other model
parameters from Popescu et al. (2011): the bulge-to-disk ratio ($B/D$), the
$SFR$ determining the luminosity of the young stellar population (thin disk),
and the parameter expressing the normalised luminosity of the old stellar
population in the disk, $old$, to derive combined radiation fields of spiral 
galaxies. This
concept, which relies on stellar light being an additive quantity,  was 
originally introduced in Tuffs et al. (2004) to
describe the attenuation of stellar light.
Here we only show radiation fields in direct stellar light (UV/optical/NIR), as
these are the ones that can be separately calculated and have additive
properties. Radiation fields in dust re-emission will be the object of a
future paper, as these cannot be separated according to the different stellar
components heating the dust, nor have they additive properties. This is because
there is no linear transformation between dust emission and absorption. For 
all these reasons, the radiation fields in dust re-emission can only be 
presented as combined radiation fields, characteristic of a galaxy as a whole.

This paper is organised as follows. The general characteristics of the
calculations are presented in Sect.~\ref{sec:simulations}. In Sect.~\ref{sec:library}  we present the
library of radiation fields and show how the energy densities should be
scaled according to the size and luminosity of a given galaxy. In
Sect.~\ref{sec:analytic_solutions} we derive analytic solutions for the 
mid-plane radiation fields of stellar disks with and without dust, which
  we use to check the accuracy of our radiative transfer calculations for the
  particular cases covered by the analytic solutions.
In Sect.~\ref{sec:spatial_variation} we
describe the spatial variation of the radiation fields for each morphological
component, referring to the analytic solutions to analyse the physical 
factors involved in shaping these variations.
In Sect.~\ref{sec:tau_variation} we correspondingly analyse the variation of the
radiation fields with face-on dust opacity.  For the specific case of bulge
radiation fields we also describe in Sect.~\ref{sec:sersic_variation} the 
change 
in energy densities when changing the S\'{e}rsic index of the 
volume stellar emissivity and apply these results to predict the corresponding
variation in heating of dust by bulge stellar populations in 
Sect.~\ref{sec:application}.
We summarise the paper in  Sect.~\ref{sec:summary}.

\section{Simulations}
\label{sec:simulations}

The simulated radiation fields are those used to generate the large
library of  dust- and  PAH-emission SEDs presented in Popescu et
al. (2011). This library is self-consistently calculated with the
corresponding library of dust attenuations first presented in Tuffs et
al.  (2004) and then in updated form in 
Popescu et al. (2011). The details of the calculations leading to the 
production of the the radiation fields are described at 
length in Popescu et al. (2011). Here we only briefly mention their main 
characteristics. All calculations were made using a modified version of 
the ray-tracing radiative transfer code of \cite{Kyl87}, which treats
  scattering utilising the method of scattered intensities (Kylafis \& Xilouris
  2005), and the dust model from Weingartner \& Draine (2001)
and Draine \& Li (2007), incorporating a mixture of silicates, graphites and 
PAH molecules.  

The radiation fields are calculated separately for the old stellar disk
(``the disk'') , the bulge 
and the young stellar disk (``the thin disk)  of a galaxy, 
all seen through a common distribution of dust. The 
geometrical model of Popescu et al. (2011) consists of both of a large-scale 
distribution of diffuse dust and stars, as well as of a clumpy component 
physically associated with the star-forming complexes. For the purpose of this 
study only the large-scale distribution of diffuse dust is considered, as it 
is this that affects the large-scale distribution of UV/optical light 
determining the values of the diffuse radiation fields.

The intrinsic volume 
stellar distributions are described by exponential functions in both radial
and vertical direction for the two stellar disks and by deprojected de Vaucouleurs
or S\'{e}rsic
functions for the bulges. The diffuse dust is distributed in two disks
associated with the old and young stellar populations, which we
refer to as the thick and thin dust disk, respectively, on account of the 
different thickness of these structures. These dust distributions are 
described by double (radial and vertical) exponential functions. The thick dust
disk is less opaque than the thin dust disk, with the relative face-on 
central dust opacity being fixed to a value of 0.387.

The length 
parameters of the model describing the volume 
emissivity for stars and dust - scale-lengths, scale-heights and 
effective radii - 
are listed in 
Table~1 in Tuffs et al. (2004). The relevant information for this work is
that the old stellar disk component has a scale-length that  decreases with 
increasing optical/NIR wavelength, as given in Table~2 in Tuffs et al. 
(2004), while the scale-height remains constant over this wavelength range. 
Similarly, the effective radius of the bulge does not vary with 
optical/NIR wavelength. The bulge is an oblate ellipsoid with an axial ratio
(thickness) of 0.6. The young 
stellar disk has a much smaller scale-height
than the older stellar disk (by a factor of 4.6), while its scale-length is
constant over wavelength and is equal to that of the old stellar disk in the B
band. 

The scale-length of the thick dust disk associated with the old stellar
population is larger (by a factor of 1.4) than that of the corresponding
stellar disk, while its scale-height is smaller (by a factor of 1.5) than the
scale-height of the old stellar disk. In contrast, the young stellar disk
spatially coincides with the associated thin dust disk (same scale-heights and
-lengths). As we will see in the next sections, the
different vertical structure of the two dust disks strongly
influences the solution for the radiation fields. The stellar and dust 
emissivities of the disks are truncated at a radius of 4.78 of the  exponential 
stellar scale-length in B-band while bulges are truncated at ten effective 
radii. The geometrical parameters of the model have either been
  empirically constrained to follow the median trends found by Xilouris et
  al. (1999) for a sample of well resolved edge-on spiral galaxies, including
  NGC891, or have been fixed from physical considerations. The physical 
interpretation of this model is described at length in Tuffs et al. (2004) and 
Popescu et al. (2011). 
A schematic representation of the geometrical model can be found in Fig.~1 in
Popescu et al. (2011).

The radiation fields are sampled on an irregular grid with sampling intervals
in radial direction ranging from 50\,pc in the centre to up to 2\,kpc in the
outer disk and with sampling intervals in vertical direction ranging from
50\,pc in the plane to up to 500\,pc in the outer halo. This scheme was
necessary to properly sample the radiation fields in the more optically thick
central regions, where the dust associated with the thin stellar disk in
conjunction with the strong positional dependence of the emissivity of the
young stellar population (in $z$) and the inner emissivity cusp of the bulge (in
both $R$ and $z$) can lead to quite small scale structures even in the diffuse
radiation fields. 

\section{The library of radiation fields}
\label{sec:library}

We have created a library of diffuse radiation fields, calculated on a
cylindrical grid $(R, z)$. The radiation fields are given for
different values of central face-on dust opacity in the B-band ${\tau_B^f}$, 
which is the only parameter of the model shaping the spatial variation of the 
RFs. Since the energy densities  of
the radiation fields are additive quantities, they scale with the
spatially integrated luminosity density at a given wavelength. Therefore we 
only calculate RFs for a fixed reference luminosity density. The radiation 
fields are separately calculated for the disk, thin disk and bulge. For the 
bulge we
considered calculations for different values of the S\'{e}rsic index of
the bulge. 

In total we performed calculations for seven values of the central
face-on dust opacity, $\tau^f_B=0.1,0.3,0.5,1.0,2.0,4.0,8.0$. Solutions 
for other values of $\tau^f_B$, such as for the case $\tau^f_B=3.5$ used to 
illustrate the solutions in this paper, can be found by interpolation. 
We considered four values for the S\'{e}rsic index of the bulge,
$n=1,2,4,8$. At each wavelength we therefore have 7 (for $\tau^f_B$) x
2 (for disk and  thin disk) + 7 (for $\tau^f_B$ ) x 4 (for S\'{e}rsic indices
of bulge) = 42 combinations at each sampled UV/optical/NIR wavelength.
We considered calculations at 15 different wavelengths, ranging from
$912\,\AA$ to $5\,{\mu}$m, as listed in Table E.2 of Popescu et
al. (2011). Thus, the library contains a total of 42 x 15= 630 files
with two-dimensional spatial grids of energy densities of the
RFs.

As mentioned before, the radiation fields have been calculated for a reference 
spatially integrated luminosity density at each wavelength and for each stellar
component. The reference luminosity is defined according to two main 
parameters of the model of Popescu et al. (2011), the star-formation rate 
$SFR$ and $old$. The $SFR$ is a parameter defining the luminosity of the 
young stellar populations and $old$ is a parameter defining the luminosity of
the old stellar population. Because the reference luminosity densities have
been defined to correspond to the unit values of the two model parameters, 
$SFR=1M_{\odot}/yr$ and $old=1$, they are referred to   
as {\it unit intrinsic spectral luminosity densities} of the young and and old
stellar populations,  $L^{young}_{\nu, unit}$ and $L^{old}_{\nu, unit}$, as
defined in Sect.~2.3 and Table~E.2 of Popescu et al. (2011). Correspondingly,
because the energy densities of the radiation fields are calculated for the 
unit $L^{young}_{\nu,  unit}$ and $L^{old}_{\nu, unit}$, they are referred to
as the {\it unit radiation fields}, as defined in Sect.~2.5.2 of 
Popescu et al. (2011): $u^{disk}_{\lambda,unit}$, $u^{tdisk}_{\lambda,unit}$
and $u^{bulge}_{\lambda,unit}$. The unit radiation fields are calculated for a 
reference disk size, as defined
in Popescu et al. (2011). The reference size  is $h_{s, ref}=5670$\,pc, defined
as the intrinsic radial scale-length of the volume stellar emissivity of the 
disk in the B-band.

Thus, the $u^{tdisk}_{\lambda,unit}$ are calculated
for a thin stellar disk having $L^{young}_{\nu,  unit}$ within the truncation
radius of $4.78\times h_{s, ref}$ (27.1\,kpc); 
the $u^{disk}_{\lambda,unit}$ are calculated for a stellar disk having
$L^{old}_{\nu, unit}$ within the truncation radius of $4.78\times h_{s, ref}$
and the $u^{bulge}_{\lambda,unit}$ are calculated for a bulge
having $L^{old}_{\nu, unit}$ within a truncation radius of 10 effective radii. 
The library of radiation fields presented in this
paper and made available at CDS is therefore given in the form of unit
radiation fields.

For morphological components of galaxies with different luminosities 
$L_{\lambda}^{disk}$,$L_{\lambda}^{tdisk}$, $L_{\lambda}^{bulge}$  and disk
sizes $h_s$, the radiation
fields at a position $(R,z)$ can be scaled according to the formula

\begin{eqnarray}
u_{\lambda}^{disk}(R,z) &  = & u_{\lambda, unit}^{disk} (R_{ref}, z_{ref}) *
\frac{L_{\lambda}^{disk}}{L_{\lambda, unit}^{old}} * \frac{h_{s, ref}^2}{h_s^2}\\
u_{\lambda}^{tdisk}(R,z) & = & u_{\lambda, unit}^{tdisk} (R_{ref}, z_{ref}) *
\frac{L_{\lambda}^{tdisk}}{L_{\lambda, unit}^{young}} * \frac{h_{s, ref}^2}{h_s^2}\\
u_{\lambda}^{bulge}(R,z) & = & u_{\lambda, unit}^{bulge} (R_{ref}, z_{ref}) *
\frac{L_{\lambda}^{bulge}}{L_{\lambda, unit}^{old}} * \frac{h_{s, ref}^2}{h_s^2}
\end{eqnarray}
where 
\begin{eqnarray}
R_{ref} & = &  R * \frac{h_{s,ref}}{h_s}\\
z_{ref} & = & z * \frac{h_{s,ref}}{h_s}
\end{eqnarray}

When combined with the solutions for the  UV/optical/submm SEDs found using
the library of Popescu et al. (2011), the amplitude of the radiation fields can
be found by scaling the unit radiation fields according to the model 
parameters $SFR$, $old$, bulge-to-disk ratio $B/D$ and the clumpiness factor
$F$, using the prescription from Sect.~2.5.2 in Popescu et al. (2011). This 
allows 
observables calculated from the radiation fields, such as gamma-rays or radio
emission, to be self-consistently combined with the solutions for the
integrated SEDs, thus expanding the range of applicability of the radiation
transfer model to a broader range of energies and physical quantities.

As noted in Popescu et al. (2011), in any dusty galaxy $h_s$ will not 
actually be a
directly observable quantity, even if the distance to the galaxy is
known.  As described and quantified in M\"ollenhoff et al. (2006) and
in Pastrav et al. (2013a), the apparent size of a galaxian disk
obtained from photometric fits to optical images will differ from the
true size due to the stronger attenuation of light at the centre of
galaxies compared to the outer regions. Furthermore, when bulge-disk
decomposition is performed on dusty galaxies, there will be additional
effects due to dust on the decomposition itself (Pastrav et al. 2013b), 
resulting in
additional differences between the intrinsic and the measured sizes of the 
disk. Thus, when scaling
radiation fields, the measured scale-length (or effective radius) of
the disk should be corrected for the effects of dust. Corrections for these
effects can be found in Pastrav et al. (2013a) and Pastrav et al. (2013b). 
Since the corrections from Pastrav et al. were derived using the same 
radiative transfer model as used for the calculation of the radiation fields, 
incorporation of these corrections in the scaling of the RFs allows for a
self-consistent analysis of the integrated SEDs, structural parameters and
radiation fields.

All the calculated radiation fields are made available in electronic
format. Their main characteristics concerning the spatial variation, as well as
the variation with dust opacity, are described in
Sects.~\ref{sec:spatial_variation} and \ref{sec:tau_variation}.

\begin{figure}
\centering
\includegraphics[scale=0.45]{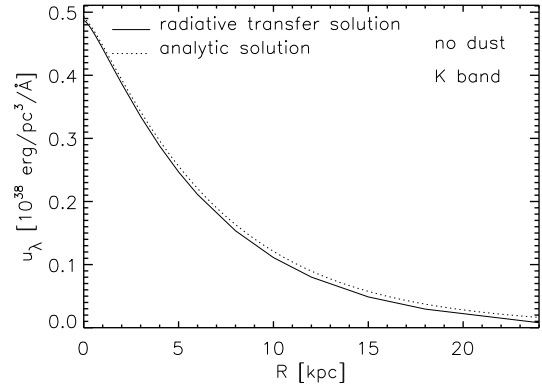}
\caption{Comparison between a radial profile of radiation fields from the 
  thin disk with no dust calculated with our radiative transfer code and the 
  one given by the analytic formula from Eq.~\ref{eq:u_z}. In both 
  cases the calculation is for the K band and
  $SFR=1\,M_{\odot}/yr$.} 
\label{fig:comp}
\end{figure}

\section{Analytic solutions}
\label{sec:analytic_solutions}

To check the accuracy of our radiative transfer calculations, but also to 
understand the trends in
the variation of the radiation fields with spatial position and dust opacity, 
we derived analytic solutions for the  energy density of the radiation fields
for limiting cases. 
Some of the formulas (those for which we give exact solutions) can also be 
used as 
benchmark cases for testing the accuracy of different radiative transfer codes 
for diffuse stellar emissivities and dust distributions. We only make 
use of the analytic solutions for comparison with the radiative transfer 
solutions in the mid-plane of the disk, even though in the case of a dustless 
disk we have derived analytic solutions at any position.

\subsection{An azimuthally symmetric stellar disk in the optically thin 
limit}

In Appendix~\ref{app:opt_thin} we derive an analytic solution for the radiation
fields in the mid-plane of an infinitely thin disk with no dust. 
This special case is
instructive, as it allows the energy density in the plane of the disk to be 
given by the sum of two terms (Eq.~\ref{eq:opt_thin}). 
The first term
(Eq.~\ref{eq:u1}) represents the contribution of all emitting
sources within a galactocentric radius $R$, while the second term 
(Eq.~\ref{eq:u2}) gives the contribution of all emitting
sources beyond the radius $R$.

\begin{figure*}
\centering
\includegraphics{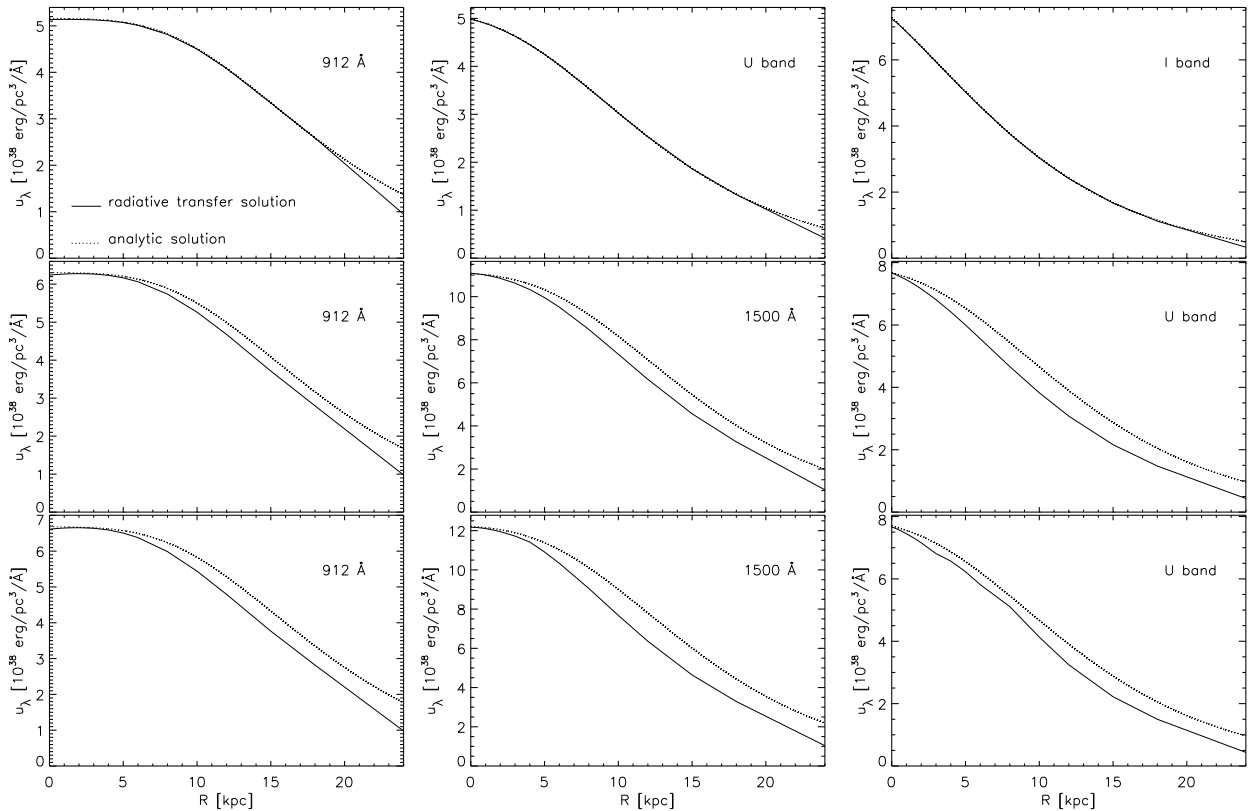}
\caption{Comparison between the solutions of the analytic formula for the
  mid-plane radiation fields of a stellar disk with one dust disk (dotted
  line) and those produced from radiative transfer calculations (solid line). 
Top: direct light only; middle: direct plus first order
scattering; bottom: total (direct plus all scattered) light. The calculations 
are 
for $\tau^f_B=3.5$ and $SFR=1\,M_{\odot}/yr$. The different panels show radial 
profiles at different wavebands.}
\label{fig:map_onedisk_td_r}
\end{figure*}

For an actual comparison of the solution of the radiative transfer 
calculation with an analytic
solution we need to also take into account the thickness of the disk in
the analytic solution, as this makes a big change. We derive in 
Appendix~\ref{app:opt_thin_z} a more general formula to take this effect into 
account (Eq.~\ref{eq:u_z}.)

We show in Fig.~\ref{fig:comp} a comparison between the radial profiles 
predicted by the radiation transfer calculation and by the analytic formula
incorporating the thin stellar disk. One can indeed see that the radiative
transfer solution almost exactly follows the prediction of the analytic 
formula (both in amplitude and shape).

\subsection{An azimuthally symmetric stellar disk with a dust disk}

In appendix \ref{app:onedust} we derive analytic solutions for the radiation
fields in the mid-plane of a stellar disk with one dust disk. The solutions are
derived for the case that the scale-height of the dust is equal to that of the
stellar disk. The calculation also assumes that the scale-height of the stellar
disk is much smaller than its scale-length. In total we derive 3 formulae: one
for direct light only, one for direct light plus first order scattered light
and one for the total direct plus scattered light.

\begin{figure*}
\centering
\includegraphics{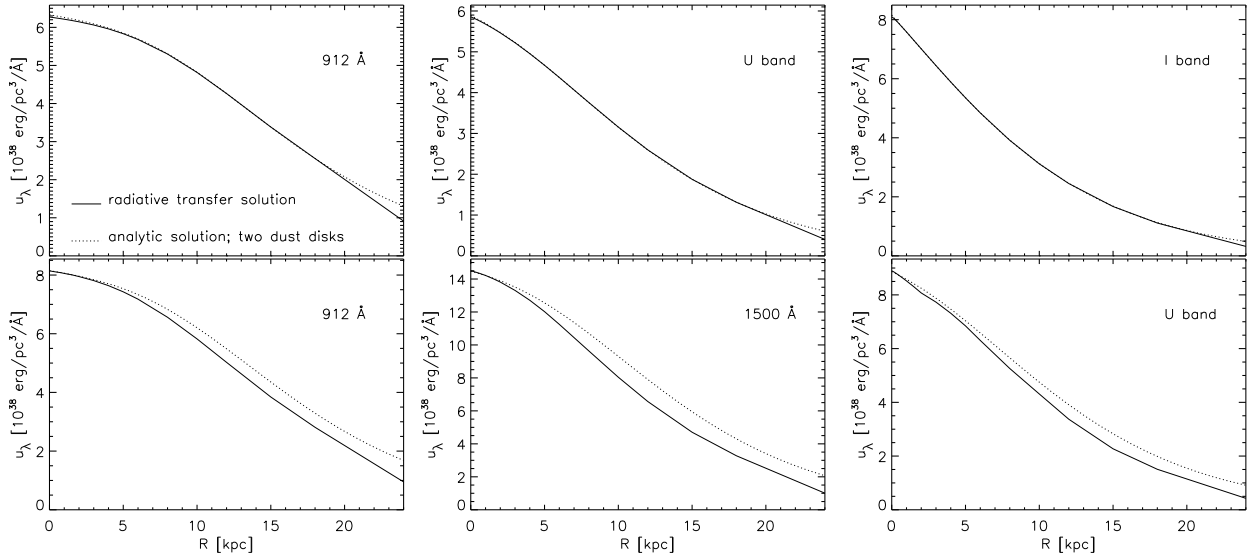}
\caption{Comparison between the solutions of the exact analytic formula for the
  mid-plane radiation fields of a stellar disk with two dust disks (dotted
  line) and those produced from radiative transfer calculations (solid line). 
Top: direct light only; bottom: total (direct plus scattered) light. The 
calculations are for $\tau^f_B=3.5$ and $SFR=1 M_{\odot}/yr$. The different 
panels show radial profiles at different wavebands.}
\label{fig:map_twodisk_td_r}
\end{figure*}

The formula for direct light is given by Eq.~\ref{eq:u3}. This is an exact
analytic formula for cases where the disk is optically thick along lines of
sight in the plane of the disk. To check the validity of this formula we
ran radiative transfer calculations for the geometry of the young stellar 
disk of our model, taking (for the purpose of this comparison) all dust 
to reside in the thin dust disk. Consequently, we took
 the scale-length of the stellar disk to be equal to that of the dust disk,
 even though
this assumption is not required by the analytic solution. In
the upper panels of Fig.~\ref{fig:map_onedisk_td_r} we show examples of
calculations done for direct light only, for
$\tau^f_B=3.5$, for three different UV/optical wavelengths, together with the
corresponding analytic solutions. One can see that the radiative transfer and
the analytic solutions agree very well across the wavelength range covered by
our calculations. The agreement is both in the radial dependence and in the
absolute level of predicted emission. The
discrepancy at larger radii is due to the inherent differences in
treating the truncation of stellar and dust emissivity in vertical and radial 
directions in the analytic and in the radiative transfer code. The fact that
the analytic formula works so well (for $\tau^f_B=3.5$) over a large range of 
wavelengths, from the UV to the NIR is due to the fact that the aspect ratio 
of our thin disk is very small, resulting in a high central edge-on dust 
opacity even in the NIR bands. It is only in the K band, in the centre of 
the profile,  that a small difference starts to
occur, which is due to the fact that the solution starts to become more 
optically thin in radial directions. Even at $\tau^f_B=0.3$ the formula gives
perfect agreement with the results of the 2D code between the wavelength range 
$912-4430\AA$. We conclude that the analytic formula for
direct light can be used as a benchmark for testing radiative transfer codes
for cases where the line of sight in the plane of the disk is optically thick.

The formula for direct plus first order scattered light is given by
Eq.~\ref{eq:final_firstorder_corr}. This formula is an approximate one, since the 
term for scattering is only exact in the extreme optically thick limit. 
Therefore it cannot be used as a benchmark case for translucent galaxies. 
Nonetheless, it can be used to qualitatively explain
 the variation of radiation fields with spatial position and dust opacity. A
 comparison between the results of our RT calculations for a one dust disk case
 for only one order scattering and those of the analytic formula are presented
 in the middle row of Fig.~\ref{fig:map_onedisk_td_r}. Since our radiative
 transfer calculations for first order scattering are exact, any departure
 between the predictions of the analytic formula and the RT profiles indicate
 the range of applicability of the formula. For very short wavelengths the
 formula predicts very well the shape 
 of the inner profiles (and $f^1_{esc}$, as defined in the appendix, is within a
 few percent) where the galaxy is very optically thick, and only 
 starts to deviate at
 larger radii. At longer UV wavelengths this deviation already happens for the
 central regions of the profiles. Nonetheless, the general trends are still 
 well explained by the analytic formula at all wavelengths.

The formula for direct plus total scattered light is given by
Eq.~\ref{eq:final_allorders_corr}. As in the case of
Eq.~\ref{eq:final_firstorder_corr}, this is also an approximation,  and
cannot be used as a benchmark. The comparison between the radial profiles
obtained from the 2D RT code and from the analytic formula (lower row of
Fig~\ref{fig:map_onedisk_td_r}) show the same qualitative behaviour as for 
the case where only first order scattering is included.

 \subsection{An azimuthally symmetric stellar disk with two dust disks}

Following the same approach as for the azimuthally symmetric stellar disk with
one dust disk, in appendix~\ref{app:twodust} we derive the corresponding
analytic formulae for the case of an azimuthally symmetric stellar disk with 
two dust disks. This is actually the case
included in our library of radiation fields, and in particular it can be used to
describe our thin disk. 

The exact formula for direct light is given by Eq.~\ref{eq:utotal_dl}. A
comparison between our 2D radiative transfer calculations and the results of
this formula is given in the top row of 
Fig.~\ref{fig:map_twodisk_td_r}, showing the same
excellent agreement as found for the single dust disk case.
The formula for the total radiation (direct plus scattered components) is given
in Eq.~\ref{eq:twodust_final_allorders_corr} and the corresponding comparison
with the numerical solution is given in
the bottom row of Fig.~\ref{fig:map_twodisk_td_r}. Although this comparison reveals a
comparably good agreement in the shape of the profile as found for the case of
the single dust disk, the $f_{esc}$ in
Eq.~\ref{eq:twodust_final_allorders_corr} takes higher values, due to
 a higher fraction
of the dust being in an optically thin configuration.

\section{Spatial variation of  radiation fields}
\label{sec:spatial_variation}

Here we describe the spatial variation of the radiation fields of each 
individual stellar component of our model. Since, as mentioned before, the 
solutions for each component are
additive, solutions for composite structures can be found by weighted addition,
according to the intrinsic stellar luminosity of each component. All examples
are for the case $\tau^f_B=3.5$. This value of dust opacity corresponds to the 
solution we found for NGC~891 (Popescu et al. 2011) and for statistical samples
of galaxies (Driver et al. 2007, Grootes et al. 2013).

\begin{figure*}
\centering
\includegraphics{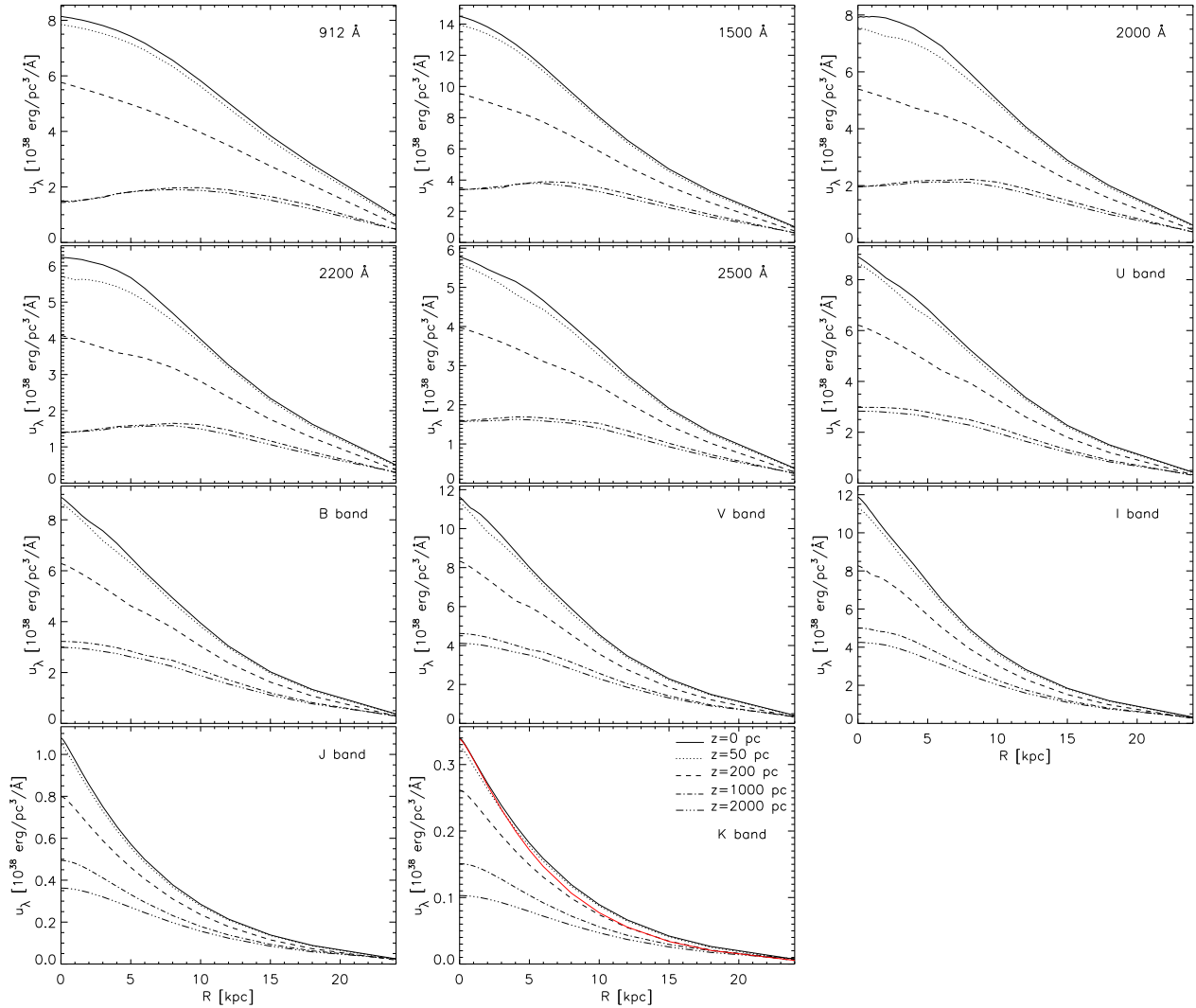}
\caption{Radial profiles of unit radiation fields from the {\bf thin disk}, for
  $\tau^f_B=3.5$.  The solid line
  represents the profiles in the plane ($z=0$\,pc) while the other
  lines (see legend) represent profiles at 
  different vertical distance from the plane ($z=$50, 200, 1000 and
  2000\,pc). The different panels show profiles at different wavebands, from
  the UV to the NIR. The profile
overplotted in red in the bottom right panel (the K-band case) represents 
the dustless radiative transfer solution scaled to the $\tau^f_B=3.5$ solution.}
\label{fig:map_radiation_fields_td_r}
\end{figure*}

\subsection {The thin disk}
\label{subsec:td}

\subsubsection{The radial profiles}\label{subsec:td_r}

The radial variation of the RF of the thin disk changes drastically between an
optically thick and optically thin case. To illustrate this, we show in 
Fig.~\ref{fig:map_radiation_fields_td_r} examples of RFs calculated at different
wavelengths. We first describe the behavior
of the RF in the plane of the disk ($z=0$). 

i) {\it The optically thin case; $z=0$}. This arises at longer wavelengths, in
the J and K bands, where the profile shows a monotonic decrease with increasing
radial distance. For comparison, the solution with no dust at all is 
overplotted with a red line in the panel corresponding to the K-band. One can see
that the radial profile calculated for the longer optical wavelengths and for 
$\tau^f_B=3.5$ tends towards the completely optically thin solution, which
shows an even steeper decrease with increasing radial distance.
This is because for  a thin
disk geometry the component of energy density arising from sources
within galactocentric radius $R$ tends (at high radii) towards an 
inverse-square law for a point source situated at $R=0$ and with luminosity 
equal to the total enclosed luminosity within $R$. This characteristic is
  explicitly seen in the formula for the exact analytic solution of an
  infinitely thin disk given in Eq.~\ref{eq:opt_thin}, \ref{eq:u1}, 
and \ref{eq:u2}. Specifically, there is an $1/R^2$ term in front of the
  integral in Eq.~\ref{eq:u1} denoting the contribution to the radiation fields
  from emitters interior to $R$. This term dominates at higher radii.
The second term in Eq.~\ref{eq:opt_thin} (as given in Eq.~\ref{eq:u2}) gives the
contribution from sources
beyond $R$, and dominates at low radii. This term gives a more slowly 
varying component for the RFs than the first term from the interior sources 
(Eq.~\ref{eq:u1}), causing the fall-off in energy
density to be shallower than an inverse square law in the inner disk. 
Beyond the radius enclosing half of the 
total luminosity, there will be a transition between the predominance of these
two terms, marking the crossover to the inverse square law behaviour.
This can also be seen in the comparison between the analytic and radiative
transfer solutions shown in Fig.~\ref{fig:comp}.

\begin{figure*}
\centering
\includegraphics{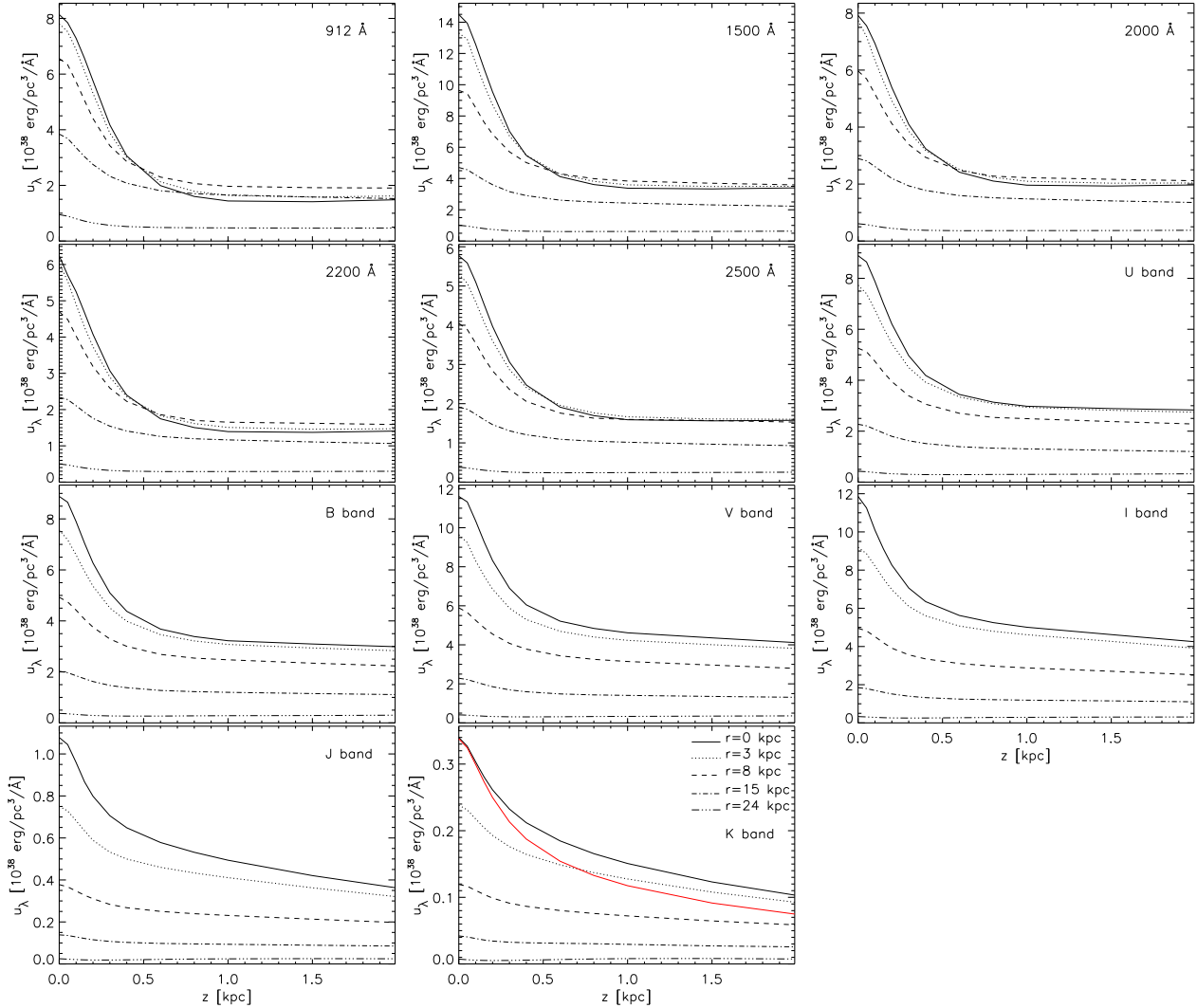}
\caption{Vertical profiles of unit radiation fields from the {\bf thin disk}, for
  $\tau^f_B=3.5$. The solid 
  line represents the profiles going through the centre ($R=0$\,pc) while the 
  other lines (see legend) represent profiles at 
  different radial distances from the plane ($R=$3, 8, 15 and
  24\,kpc). The different panels show profiles at different wavebands, from
  the UV to the NIR. The profiles at different wavebands have been scaled 
as in  Fig.~\ref{fig:map_radiation_fields_td_r}.The profile
overplotted with the red line in the bottom right panel (the K-band case) represents 
the dustless solution. }
\label{fig:map_radiation_fields_td_z}
\end{figure*}

ii) {\it The optically thick case; $z=0$.} This arises at shorter UV
wavelengths, where one can see a flat profile in the inner region, followed by
a monotonic decrease with increasing radial distance. The inner flat profile is
a direct consequence of the optically thick character of the solution. Thus,  
in an optically thick case the
radiation seen at each point is proportional to the stellar emissivity within
the horizon sphere\footnote We define the term ``horizon surface''  to
  denote the locus of positions for which the 
line integral of opacity from a point within an optically thick body is
unity. In the extreme optically thick case the horizon surface is a sphere,
which we call ``horizon sphere''.
divided by the dust opacity:
\begin{eqnarray}\label{eq:opt_thick}
u(R)= \frac{\eta(R)}{\kappa(R)}\frac{1}{\left(1-A\right)\,c}
\end{eqnarray}
where $\eta$ is the local stellar emissivity, $\kappa$ is the local extinction
coefficient (defined as the probability of extinction of a photon per unit 
length), $A$ is the albedo of the dust grains and $c$ is the speed of light.  
The horizon sphere is then the sphere with radius equal to
$1/\kappa$ (i.e. the typical local mean free path of photons). 
Eq.~\ref{eq:opt_thick} is a generalised version of the specific solutions for 
the single and two dust disks given in Eqs.~\ref{eq:main} and 
\ref{eq:twodust_final_allorders_corr_optthick}. Eq.~\ref{eq:opt_thick} is
always valid, irrespective of geometry, for very optically thick cases. In the 
specific case considered here, Eq.~\ref{eq:opt_thick} refers only to the local 
mid-plane volume emissivity and opacity. Thus, it will give the
correct solution in all cases where the horizon is
so close that any spatial variations in emissivity and opacity within
the horizon can be ignored. The subtlety is that, when the horizon viewed
from the mid-plane gets larger (e.g. when moving in radius towards the outer
disk) Eq.~\ref{eq:opt_thick} still gives the correct solution as long as the 
variation (primarily in z) of stellar emissivity and opacity is the same. 
This is the case when the horizon sphere is contained within the thin
disk, since both the stellar emissivity
and the dust opacity have the same scale-height, and thus, the same spatial 
variation within the horizon sphere. Then the mid-plane RF will remain 
constant. 

As the radial distance increases, the horizon surface starts to become a prolate
ellipsoid, because light
can propagate to larger distances vertically than in the 
radial direction before reaching the horizon (the scale-height of the 
dust is much shorter than its scalelength, by a factor 0.016 in our model).
When the horizon gets so big that new geometrical components
in emissivity and/or opacity enter into the horizon ellipsoid, and these new
geometrical components have a different 
ratio $\eta/\kappa$ than the ratio at $z=0$, Eq.~\ref{eq:opt_thick} will 
break down. This happens beyond a certain galactocentric radius, where the 
height of the horizon ellipsoid becomes larger than the scale-height of the 
thin disk. Then the spatial variation of total 
opacity in $z$ within the horizon ellipsoid is no longer well approximated by 
the  spatial variation in opacity of dust belonging to the thin disk. This is
due to the dust opacity from the thick dust disk (which has a larger scale 
height than that of the stellar emissivity in the thin disk).
Thus, according to Eq.~\ref{eq:twodust_final_allorders_corr}, the RFs 
will start to fall off from the maximum value reached in the (completely) 
optically thick case 
(Eq.~\ref{eq:twodust_final_allorders_corr_optthick}), explaining the 
reduction in the mid-plane RFs in the outer disk. Eventually the RFs will 
tend towards an inverse square law decrease, which sets in at radii beyond 
which all outwards lines of sight (in the plane as well as perpendicular to 
the plane) become optically thin.

iii) {\it Intermediate cases; $z=0$}. At longer UV/shorter optical wavelengths,
  the profiles show a mixed behaviour, between the optically thin and optically
  thick limits. Essentially they resemble the profiles from the optically
  thick cases, with the difference that the flat part in the
  inner radius is missing. This is due to the fact that already in the inner
  region the horizon surface is given by a prolate ellipsoid and not by a
  sphere. 
  Because the height of this ellipsoid is always significantly
  larger than the scale-height of the thin stellar disk and increases 
  progressively with increasing radius, there is a corresponding decrease
    in the ratio of stellar emissivity to dust opacity (in vertical 
 direction), resulting in the monotonic decrease of  the RFs.

Secondly, we describe the behaviour of the radial profiles at various vertical 
distances from the plane.

i) {\it The optically thin case; high z.} As before, this arises at longer 
optical wavelengths. 
The profiles show a similar behavior to the radial profiles in the plane of the
disk, but with a shallower decrease. This happens because the fractional 
change in the distances to all emitting elements is smaller from an
outside view than from an inside view (within the disk).

ii) {The optically thick case; high z.} The profiles at large
vertical distance are rather flat, as seen in
Fig.~\ref{fig:map_radiation_fields_td_r} in the UV range at $z=1-2$\,kpc.
This qualitative change in the solution for high $z$ is because the thick dust 
disk, rather than the thin dust disk, controls the emission escaping from the 
thin disk. Even from a vantage point 2\,kpc above the disk (the maximum height 
calculated here), the thin stellar disk is viewed through an optically thick 
layer of dust in the thick disk extending over almost $2\,\pi$ sterad.  The 
level of 
the energy densities of the RF is determined by the ratio of thin disk 
emissivity to thick disk $\kappa$ at the depth of the emitting surface
\footnote{ When viewing any optically thick emitting body from the outside,
 the photons
that one sees from that body will on average have originated from positions
with line-of-sight optical depths to the observer of order unity. For bodies 
with a strong line of sight gradient in optical depth these 
positions at which the escaping photons are emitted will be at a rather 
localised distance from the observer, thus lying on a surface. We call this
surface the ``emitting surface''. The characteristics of the
radiation from the emitting surface will be determined by the characteristics
of the radiation field in the optically thick layer just below the emission 
surface, which, in turn, will be determined by the local ratio 
${\eta}/{\kappa}$ just below the emission surface.} 
 from which the escaping photons are emitted. Because of
the much steeper z dependence of the thin disk emissivity compared to that of 
the thick dust disk opacity, this depth is well below the $\tau=1$ surface 
(as seen from above)- ie the effective line-of-sight optical depth to the 
emitting surface is $> 1$.
This in turn means that the emitting surface has a larger extent than
the extent of the $\tau=1$ surface as seen from $z=0$, explaining the much 
flatter form of the RF curve with radius for the $z=2$\,kpc case compared to the
$z=0$\,kpc case.

\begin{figure*}
\centering
\includegraphics{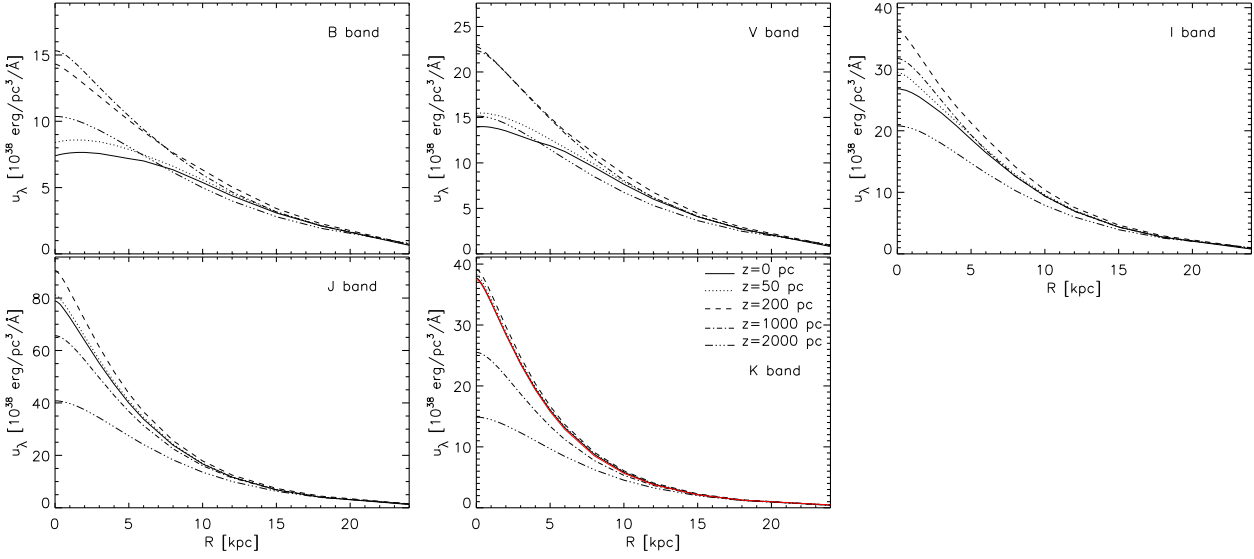}
\caption{Radial profiles of unit radiation fields from the {\bf disk}, for $\tau^f_B=3.5$.
  The solid line
  represents the profiles in the plane ($z=0$\,pc) while the other
  lines (see legend) represent profiles at 
  different vertical distance from the plane ($z=$50, 200, 1000 and
  2000\,pc). The different panels show profiles at different wavebands, from
  the B band to the NIR. The profile
overplotted in red in the bottom right panel (the K-band case) represents 
the dustless solution scaled to the $\tau^f_B=3.5$ solution.}
\label{fig:map_radiation_fields_d_r}
\end{figure*}

\subsubsection{The vertical profiles}\label{subsec:td_z}

The trends in the vertical profiles can be easily explained following the
trends already discussed for the radial profiles. Thus, at shorter wavelengths,
where the disks are optically thick, the profiles in the centre ($R=0$) show a 
decrease
in the energy density with increasing $z$, until a certain distance above the
disk, from where the RF remain fairly constant (see  
Fig.~\ref{fig:map_radiation_fields_td_z}). As discussed for the radial
profiles, in an optically thick case, the RF vary according to
Eq.~\ref{eq:opt_thick}, as long as the vantage point is within the dust layer. 
 However, unlike the radial variation, in the vertical direction the RFs are
 affected by the two dust disk structure, where the disk opacity has a
 different variation from that of the stellar emissivity. The RFs will 
therefore 
decrease with increasing vertical distance from the disk, and this behavior
continues also for vertical profiles at larger radii, since the disk remains
optically thick for a large range of radial distances. 
Moving still further out in $z$, it becomes possible for light to freely 
escape upwards from an  emitting surface (below) which marks the transition 
(in the vertical direction) from  optically thick to optically
thin. At a vantage point more than 500\,pc up, the contribution to the RFs
from this surface dominates the in situ contribution
from the thin stellar disk, whose emissivity is tiny at more than a few scale 
heights. Because at short wavelengths the thin disk is optically thick (face-on)
over a large range of galactocentric radii, the emitting surface extends over 
almost $2\pi$ sterad. Thus, a vantage point at larger vertical distances above 
the disk will essentially see 
a constant surface-brightness until large radii, causing the RFs to remain
constant higher-up above the disk.

The trend of decreasing the RF with increasing $z$ followed by a flat trend 
with $z$ is gradually changing from shorter to longer wavelengths, as the disk
becomes more optically thin. Eventually in the optically thin case there is
only a continuous decrease of the RF with increasing $z$. For comparison, 
the solution with no dust at all is overplotted with a red line in the panel 
corresponding to the K-band. One can see that the vertical profile calculated
for the longer optical wavelengths and for $\tau^f_B=3.5$ tends towards the
completely optically thin solution, though the latter shows a much  
steeper decrease with increasing vertical distance.

\begin{figure*}
\centering
\includegraphics{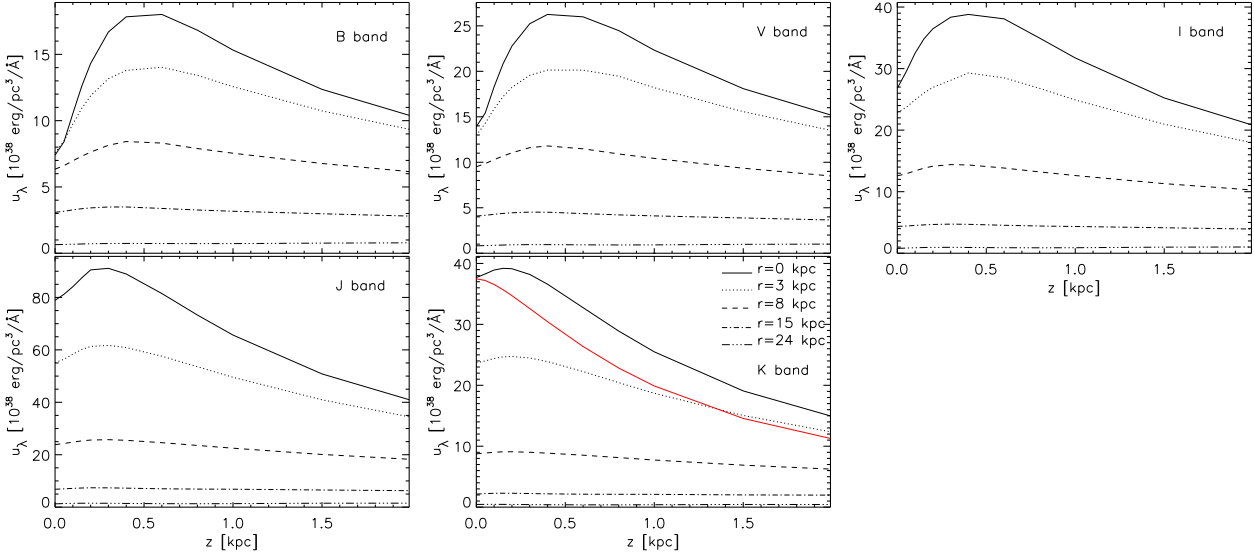}
\caption{Vertical profiles of unit radiation fields from the {\bf disk}, for
  $\tau^f_B=3.5$. The solid 
  line represents the profiles going through the centre ($R=0$\,pc) while the 
  other lines (see legend) represent profiles at 
  different radial distance from the plane ($R=$3, 8, 15 and
  24\,kpc). The different panels show profiles at different wavebands, from
  the B band to the NIR. The profile
overplotted in red in the bottom right panel (the K-band case) represents 
the dustless solution.}
\label{fig:map_radiation_fields_d_z}
\end{figure*}

Coming back to the optical thick cases, but looking at the profiles at larger 
radii (Fig.~\ref{fig:map_radiation_fields_td_z}) one can see a similar behavior
to that in the centre of the galaxy. However, at larger vertical distances
above the disk, where the energy densities become constant, the overall level
of the constant is higher than that of the RF in the centre of the galaxy. So
the profiles do not show a monotonic behavior in this respect. This is because,
although the vantage point above the disk will also see a constant surface
brightness, the level of this is mainly determined by the thin (small
scale-height) dust disk, 
since the thick dust disk (with the larger scale-height) becomes more optically
thin. In other words at smaller radii and high z the radiation fields are 
equal to 
\begin{eqnarray}\label{eq:opt_thick_two}
u(R)= \frac{\eta(R)}{\kappa_1(R)+\kappa_2(R)}\,\frac{1}{\left(1-A\right)\,c}
\end{eqnarray}
while at intermediate radii and high z the radiation fields are equal
to 
\begin{eqnarray}\label{eq:opt_thick2}
u(R)= \frac{\eta(R)}{\kappa_2(R)}\,\frac{1}{\left(1-A\right)\,c}
\end{eqnarray}
where $\kappa_1$ and $\kappa_2$ are the local opacities of the thick and 
thin dust disks, respectively. These are generalisations of
Eq.~\ref{eq:twodust_final_allorders_corr_optthick}, where the mid-plane
opacities have been replaced by local opacities. 
As a consequence, the surface brightness increases at a intermediate radii,
since the layer of constant surface brightness moves towards the plane of the disk. There will be a radius at which a maximum in the
value of the energy density of the RF at larger $z$ is attained, after
which this will start to slowly decrease again. This causes the non-monotonic
behaviour of the radiation fields above the disk with increasing radius. At
longer wavelength this never happens, because the thin dust disk becomes
transparent even in the inner radii.

\subsection{The Disk}\label{subsec:d}

In our model we consider the disk to be a source of only optical/NIR photons
from an older stellar population, with stars old enough to have been
dynamically heated to a thicker disk configuration.

\begin{figure*}
\centering
\includegraphics{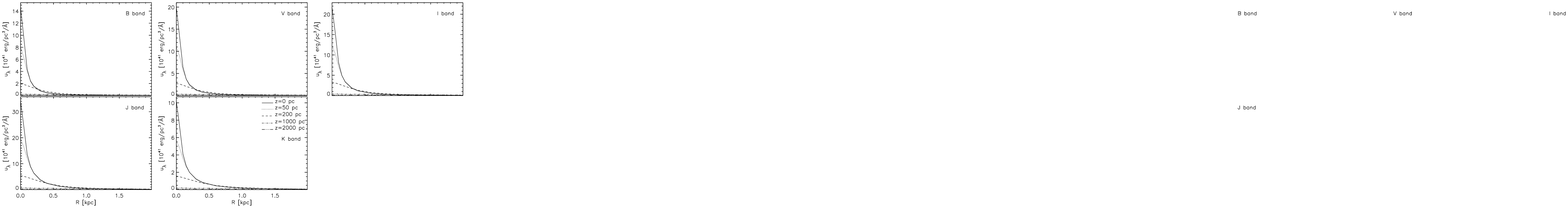}
\caption{Radial profiles of unit radiation fields from the {\bf bulge}, for
  $\tau^f_B=3.5$. The solid line
  represents the profiles in the plane ($z=0$\,pc) while the other
  lines (see legend) represent profiles at 
  different vertical distances from the plane ($z=$50, 200, 1000 and
  2000\,pc). The different panels show profiles at different wavebands, from
  the B-band to the NIR. The calculations are for de
  Vaucouleurs bulges.}
\label{fig:map_radiation_fields_b_r}
\end{figure*}
\begin{figure*}
\centering
\includegraphics{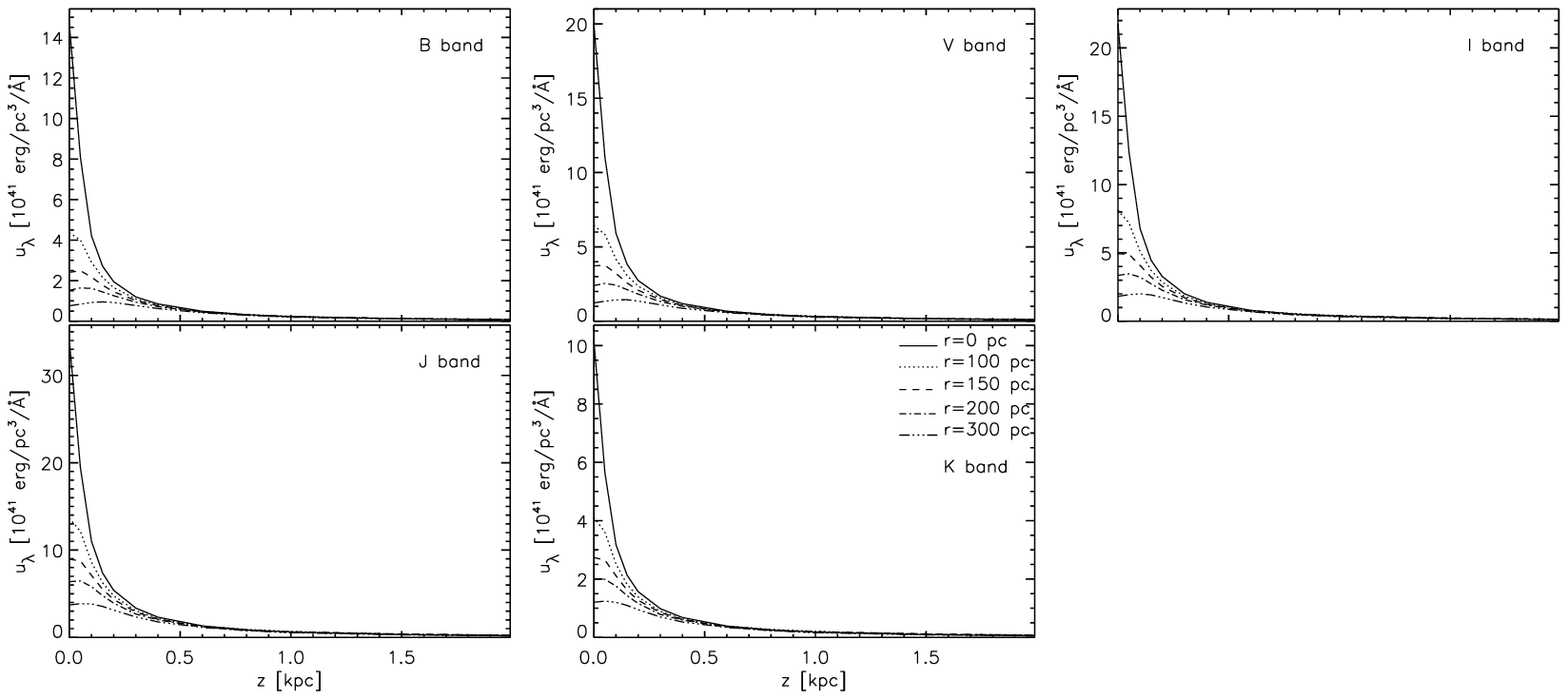}
\caption{Vertical profiles of unit radiation fields from the {\bf bulge}, for
  $\tau^f_B=3.5$. The solid 
  line represents the profiles going through the centre ($R=0$\,pc) while the 
  other lines (see legend) represent profiles at 
  different radial distance from the plane ($R=$3, 8, 15 and
  24\,kpc). The different panels show profiles at different wavebands, from
  the B-band to the NIR. The calculations are 
  for de Vaucouleurs bulges.}
\label{fig:map_radiation_fields_b_z}
\end{figure*}

At shorter optical wavelengths the radial profiles at $z=0$ are rather flat 
(see Fig.~\ref{fig:map_radiation_fields_d_r}). This behaviour is quite different
from that of the thin disk at the same wavelengths (see
Fig.~\ref{fig:map_radiation_fields_td_r}) and appears because the stellar 
disk has a larger scale-height than that of the thin stellar disk.  Thus, as
in the thin disk case, at small radii and at $z=0$ the disk is optically thick 
in the radial direction and more optically thin in the vertical direction. 
Thus the horizon surface is given by a prolate ellipsoid.
Although the height of this ellipsoid continuously increases with radius
(as in the case of the thin disk), this still remains smaller than the 
scale-height of the stellar disk. So, unlike the thin disk case at shorter
optical wavelength (e.g. B band), the ratio
between stellar emissivity to dust opacity will remain fairly constant within
the galactocentric radius for which the height of the ellipsoid is smaller than
the scale-height of the stellar disk. Therefore, radial profiles for the 
radiation fields are flatter in the inner disk.

At larger radii
though, the disk become optically thin, consistent with a radial profile of RF
decreasing with increasing radius. At longer optical/NIR wavelengths, the disk
becomes optically thin even in the inner part, therefore the RF will show a
continuous decrease with increasing radius. For comparison, 
the solution with no dust at all is overplotted with a red line in the panel 
for the K-band. One can see that the radial profile calculated
for the longer optical wavelengths and for $\tau^f_B=3.5$ is very well
described by the completely optically thin solution.

\begin{figure*}
\centering
\includegraphics{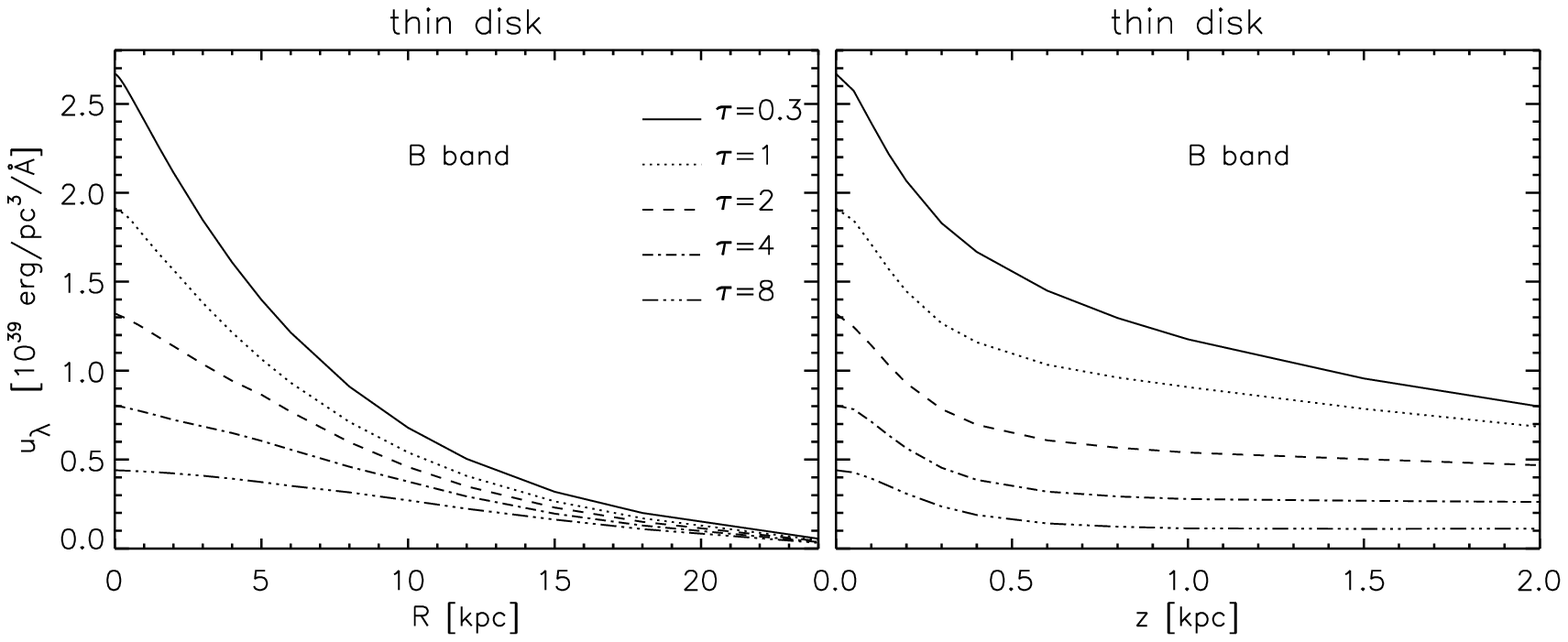}
\caption{Left: B band radial profiles of unit radiation fields 
from the {\bf thin disk} calculated in 
the plane ($z=0$\,pc). Right: Corresponding vertical profiles calculated 
through the centre ($R=0\,$pc). The different lines represent profiles
corresponding to 
different central face-on dust opacities in the B-band $\tau^{f}_{B}$ (see
legend).}
\label{fig:map_radiation_fields_td_tau}
\end{figure*}

The behavior of the RFs above the disk can be better seen in the vertical profiles (see Fig.~\ref{fig:map_radiation_fields_d_z}). The
dominant feature of these profiles is the existence of a maximum in the energy
density of the RFs at an intermediate distance above the plane. Thus, the RFs 
increase from $z=0$ until a certain height, and then they decrease again.
This is because the stellar emissivity in the disk has a larger scaleheight
than the scaleheight of both dust disks. So in the 
optically thick regime the energy density of the RFs goes up until the
solution for the 
optically thin
case is reached. The height for which the
maximum is obtained decreases continuously with increasing optical
wavelength (due to the decrease in the optical thickness
of the disk). In the K-band the maximum is
obtained at very small distances above the plane, but it never reaches the
$z=0$ plane. This shows that the completely optically thin solution is
  approached, but still not reached. The dustless case (the red
  profile in Fig.~\ref{fig:map_radiation_fields_d_z}) indeed shows a maximum at $z=0$.

At larger radii the vertical profiles are very flat in all cases, because the
typical distance to the emitting regions, irrespective of whether they are 
optically thin or optically thick, does not change much. 

\subsection{The Bulge}
\label{subsec:b}

Here we discuss the spatial trends of radiation fields for de
Vaucouleurs bulges (corresponding to a S\'{e}rsic index
$n=4$). Deviations from these trends in bulges with different S\'{e}rsic
indices will be  discussed in Sect.~\ref{sec:sersic_variation}.
Unlike the disk, the RFs of the bulge in the plane of the disk ($z=0$) and for
the optically thick cases (e.g. B band) do not 
show a constant trend with radial distance, even for the most optically thick 
cases (see Fig.~\ref{fig:map_radiation_fields_b_r}). On the contrary, the radial
profiles show a very steep decrease at small radii. This is because of the 
steep  dependence of stellar emissivity on radius, always ensuring a monotonic 
decrease in the ratio of stellar emissivity to dust opacity. By the same token,
the same trend can be seen in the vertical profiles at $R=0$ 
(see Fig.~\ref{fig:map_radiation_fields_b_z}).

However, going away from the centre, the vertical profiles start to
qualitatively resemble the behaviour of the disk (a maximum in the energy
densities at an intermediate distance above the plane). This is because the 
bulge has, like the disk, a layer of stellar emissivity going above the dust
layer, namely an effective radius that is larger than the scale-height of the 
dust disks. Consequently, there is also a non-monotonic behavior in the radial 
profiles at larger vertical distances, as in the case of the disk.
At larger $R$ and $z$ the RF tend to an inverse square law.

\section{Variation of radiation fields with central 
face-on dust  opacity $\tau^{f}_{B}$}
\label{sec:tau_variation}

\subsection{The Thin Disk}
\label{subsec_tau_td}

The variation of the radiation fields of the thin disk with $\tau^{f}_{B}$ is 
exemplified in Fig.~\ref{fig:map_radiation_fields_td_tau}, for radial 
positions in the plane of the disk as well as for vertical positions at
$R=0$, for a fixed wavelength (B-band). The first 
characteristic of the curves is the
decrease of the energy density of the RF with increasing  $\tau^{f}_{B}$. 
This is to be expected, as
photons coming from a wider range of distances can reach any given vantage
point in the galaxy in the optically thin case than in the
optically thick case.

The radial profiles (left panel Fig.~\ref{fig:map_radiation_fields_td_tau}) 
also show the expected trend with optical thickness, 
namely the change from a continuously decreasing energy density of the RF with 
increasing radius for 
optically thin cases, to, for optically thick cases, a profile which is almost 
flat in the inner regions, only markedly decreasing at high increasing radii.
(see detailed explanations in Subsect.~\ref{subsec:td_r}). This is similar 
with the trend with wavelength for the same spatial locations 
(see Fig.~\ref{fig:map_radiation_fields_td_r}), since the increase
in wavelength is followed by a decrease in the dust opacity at that wavelength.

The vertical profiles (right panel Fig.~\ref{fig:map_radiation_fields_td_tau}) 
show a transition from a continuously decreasing energy 
density of the RF with increasing vertical distance from the plane in the
optically thin case, to, for optically thick cases, profiles which also 
decrease at  smaller distances above the plane, but become flat at higher 
vertical positions.
This is the expected trend when moving from an optically thin to an optically
thick case, as explained in Subsect.~\ref{subsec:td_z} (see also the similar
trend with wavelength in Fig.~\ref{fig:map_radiation_fields_td_z}).

\subsection{The disk}
\label{subsec:tau_d}

\begin{figure*}
\centering
\includegraphics{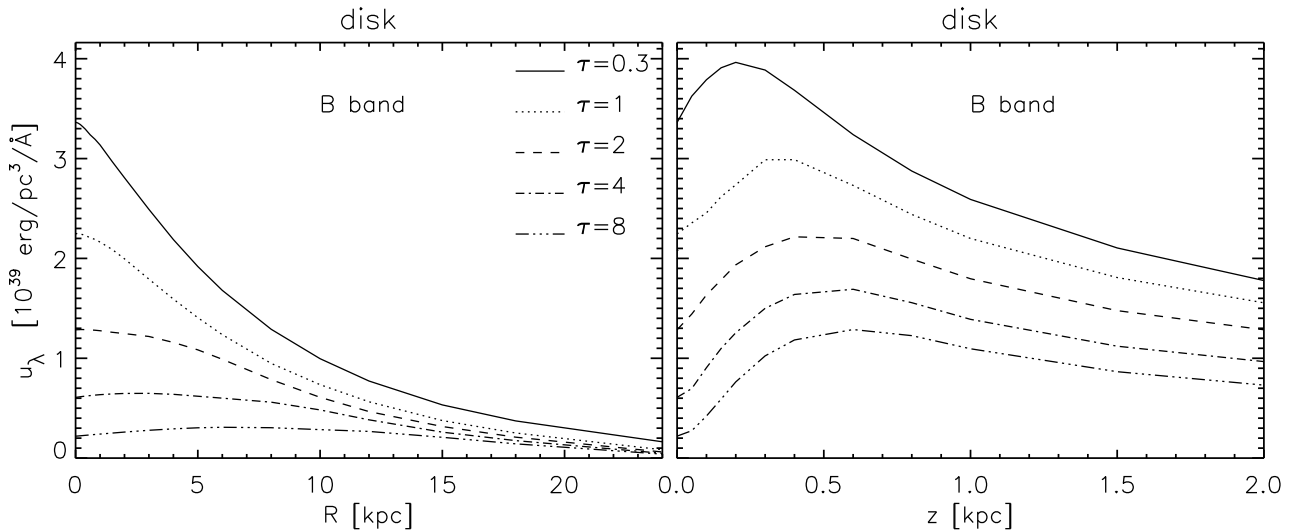}
\caption{Left: B band radial profiles of unit radiation fields 
from the {\bf disk} calculated in the 
plane ($z=0$\,pc). Right: Corresponding vertical profiles calculated through the
centre ($R=0\,$pc). The different lines represent profiles corresponding to
different central face-on dust opacities in the B-band $\tau^{f}_{B}$ (see
legend).}
\label{fig:map_radiation_fields_d_tau}
\end{figure*}
\begin{figure*}
\centering
\includegraphics{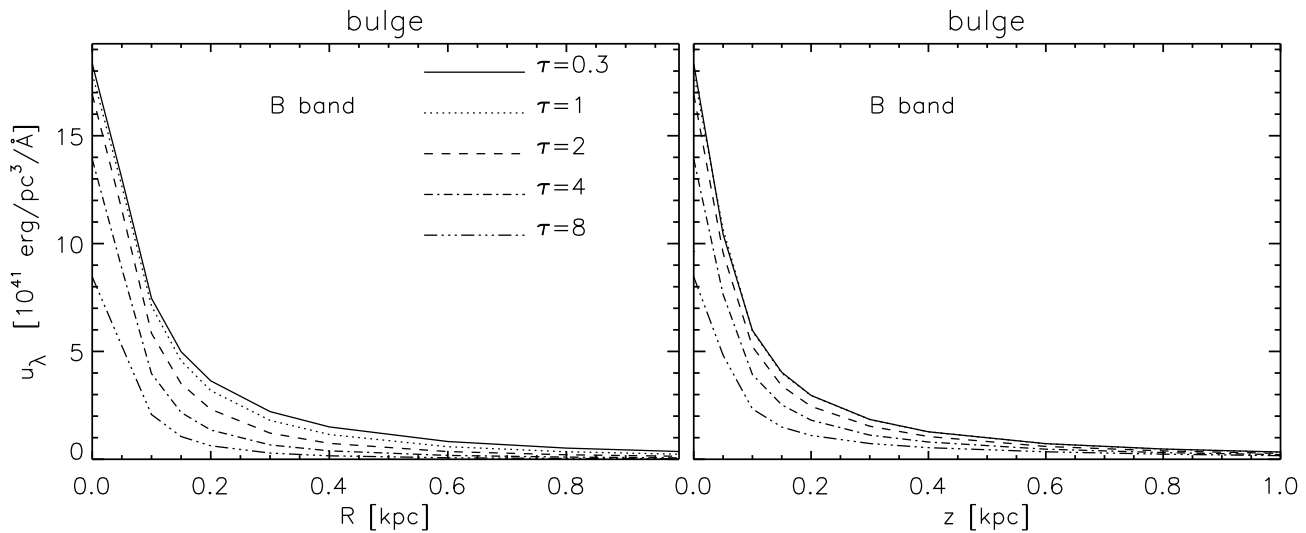}
\caption{Left: B band radial profiles of unit radiation fields 
from the {\bf bulge} calculated in the 
plane ($z=0$\,pc). Right: Corresponding vertical profiles 
calculated through the
centre ($R=0\,$pc). The different lines represent profiles corresponding to
different central face-on dust opacities in the B-band $\tau^{f}_{B}$ (see
legend). The
calculations are for de Vaucouleurs bulges.}
\label{fig:map_radiation_fields_b_tau}
\end{figure*}

As in the case of the thin disk, the radial profiles in the plane of the disk
and the vertical profiles in the centre of the disk
(given for the B-band in Fig.~\ref{fig:map_radiation_fields_d_tau}) show a 
decrease in the energy density of the radiation fields with increasing dust 
opacity $\tau^{f}_{B}$. 

The radial profiles for
the optically thin cases show a continuous decrease with increasing radius 
 (left panel Fig.~\ref{fig:map_radiation_fields_d_tau}),
since the disk is in these cases optically thin even in the inner regions. In
the optically thick cases the radial profiles are flat in the inner
radii, or even slightly increase, followed by a decrease of the energy 
densities at larger radii. The disk
remain essentially flat for radial distances for which the disk is still
optically thick. This means that the higher the face-on optical depth
$\tau^{f}_{B}$, the larger the radius for which the disk is optically
thick, and therefore the profiles are flat.

As discussed in Sec.~\ref{subsec:d}, the vertical profiles 
(right panel Fig.~\ref{fig:map_radiation_fields_d_tau})  show a maximum in the 
energy density at intermediate vertical distances, due to the fact that the
stellar emissivity in the disk has a larger scaleheight than the scaleheight of
the two dust disks. The vertical distance at which this maximum occurs 
increases with increasing dust opacity $\tau^{f}_{B}$, 
since the optical thick regime extends to higher vertical distances in an
optical thick case. This effect is particularly strong for the $\tau^{f}_{B}=8$
case, where the energy densities increase by a factor of $\sim6$, when moving
from the centre to a position 600\,pc above the plane.

\subsection{The Bulge}
\label{subsec:tau_b}

The radiation fields of the bulge also decrease with increasing
$\tau^{f}_{B}$, albeit at a much smaller rate that for the thin disk and disk 
(see Fig.~\ref{fig:map_radiation_fields_b_tau}). The
radial profiles of RFs in the plane of the disk show a monotonic
decrease with increasing radial distance, irrespective of
$\tau^{f}_{B}$. Similarly, the vertical profiles of RFs at $R=0$\,pc
show a monotonic decrease with increasing vertical distance for all
dust opacities considered here. As explained in Sect.~\ref{subsec:b},
 even in the very optically thick cases the
 profiles show a steep decrease with $R$ and $z$ in the central
 regions due to the very steep dependence of the stellar emissivity
 with radial and vertical distance.

\section{Variation of radiation fields with 
bulge S\'{e}rsic index}
\label{sec:sersic_variation}

\begin{figure*}
\centering
\includegraphics[scale=1.0]{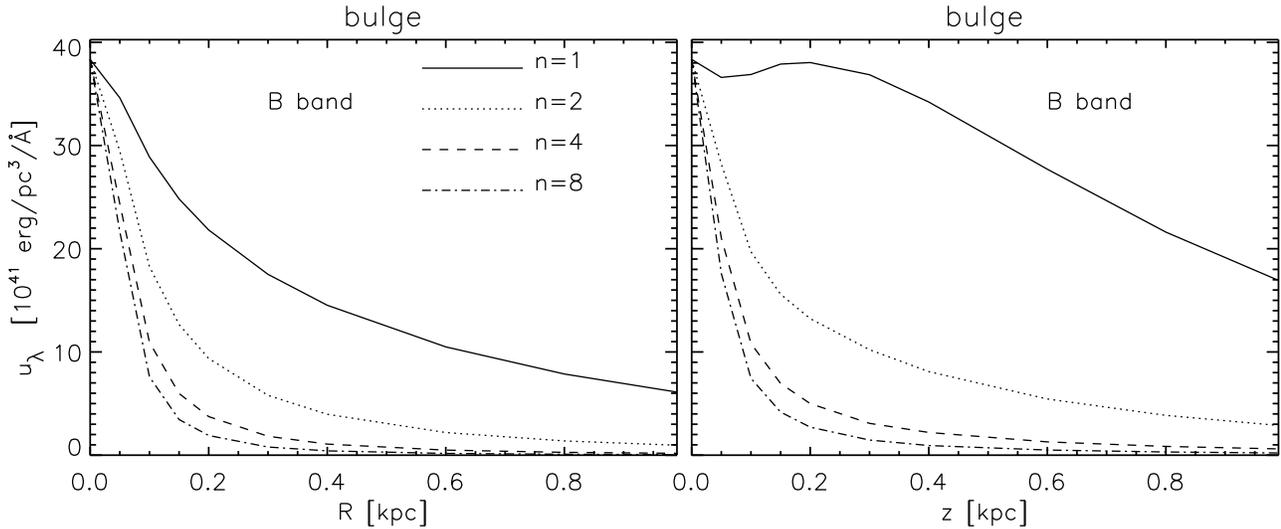}
\caption{Left: B band radial profiles of unit radiation fields from the 
{\bf bulge} calculated in the 
plane ($z=0$\,pc). Right: B band vertical profiles from the bulge 
calculated through the
centre ($R=0\,$pc). The different lines represent profiles corresponding to
different S\'{e}rsic indices of the stellar emissivity (see
legend). The profiles are scaled such that the maximum of the energy 
density of the radiation fields of a model galaxy reaches the value of 
$38.34\times10^{41}$, which is the maximum intensity for the model with $n=8$. 
The calculations are for $\tau^{f}_{B}=4.0$.}
\label{fig:map_radiation_fields_b_sersic1}
\end{figure*}

In Sects.~\ref{subsec:b} and \ref{subsec:tau_b} we discussed the
spatial variation of the radiation fields from bulges, as well as the
variation with wavelength and central face-on dust opacity. This was done for the
concrete example of de Vaucouleurs bulges. Here we investigate
whether the trends seen in the RF for bulges change when different
S\'{e}rsic indices are considered. In
Fig.~\ref{fig:map_radiation_fields_b_sersic1} we show radial and
vertical profiles of RFs at $z=0$ and $R=0$ respectively, for four
different values of S\'{e}rsic indices, $n= 1,2,4,8$,  
for the case $\tau^f_B=3.5$, and for the B-band.

The radial profiles show a monotonic decrease with increasing radial
distance, irrespective of the S\'{e}rsic index. However, for higher
S\'{e}rsic indices the decrease is steeper, in particular in the very
central regions, as expected due to the steeper dependence of the
stellar emissivity. For $n=1$ the profiles start to resemble those
from the disk, due to the exponential form of the stellar emissivity.

The vertical profiles show more pronounced differences in the trends
for the different S\'{e}rsic indices. Thus, for lower S\'{e}rsic index the
profiles become less steep. In the extreme case of $n=1$ the profile
not only flattens, but even shows a second maximum. This exponential
form of the stellar emissivity produces a solution closer to the disk
solution, where a maximum occurs at a certain vertical distance from
the disk. This is due to the fact that the stellar emissivity has a larger
scale-height than that of the two dust disks. However, unlike the
disk, there is another maximum in the plane of the disk. This is
because, although the stellar emissivity is described by an
exponential distribution, this emissivity is spatially distributed in
an ellipsoid rather than a disk.

Overall we can conclude that the radiation fields from bulges strongly depend
on the assumed S\'{e}rsic index of the bulge. Because of this we
include in
the library of radiation fields for bulges the calculations for all four values of
S\'{e}rsic indices shown above.

\section{Application: the effect of different bulge S{\'e}rsic 
indices on dust heating}
\label{sec:application}

One important conclusion of this study is that the radiation fields
from bulges strongly depend on the value of the S\'{e}rsic index of the 
stellar emissivity (see Sect.~\ref{sec:sersic_variation}). This raises the
possibility that the heating of dust in the disk may also be strongly 
influenced by the type of bulge a galaxy would host.

\begin{figure}
\includegraphics[scale=0.5]{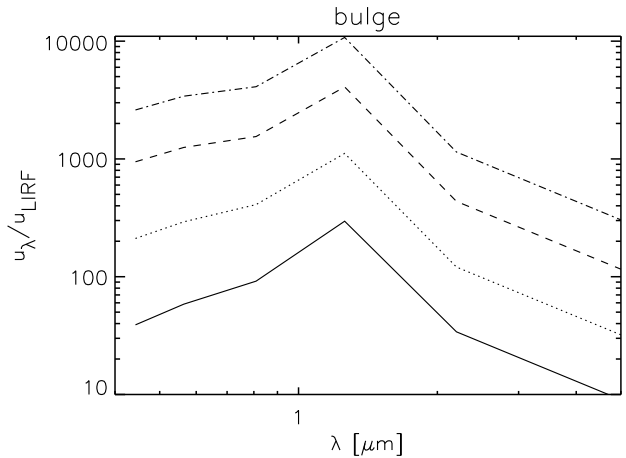}
\caption{The SED of the unit radiation fields (normalised to the LIRF) 
at $R=0$\,kpc and $z=0$\,kpc, for bulges having different S\'{e}rsic indices 
of the stellar emissivity (plotted with different lines; legend as in
Fig.~\ref{fig:map_radiation_fields_b_sersic1})}
\label{fig:SED}
\end{figure}

Indeed, when plotting the SED of the radiation fields (normalised to the local
interstellar radiation fields - LIRF - Mathis et al. 1983), in the centre of 
the bulge (Fig.~\ref{fig:SED}), one can see a substantial increase in the energy
densities from an $n=1$ case to an $n=8$ case. Without any explicit
calculation of the dust heating, this already indicates that the 
temperature of the dust heated by radiation fields coming from bulges with
different S\'{e}rsic indices will change significantly.

\begin{figure}
\includegraphics[scale=0.5]{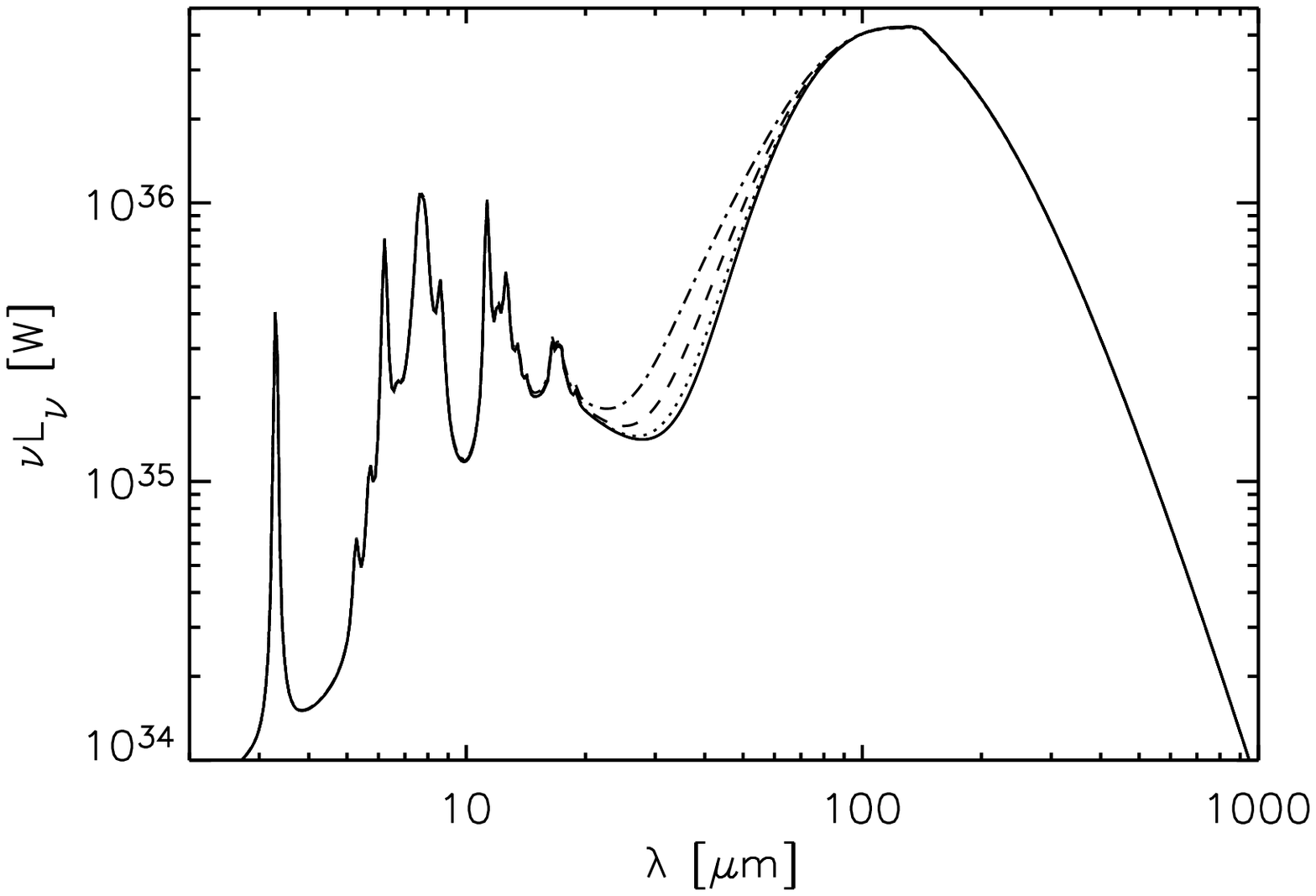}
\includegraphics[scale=0.5]{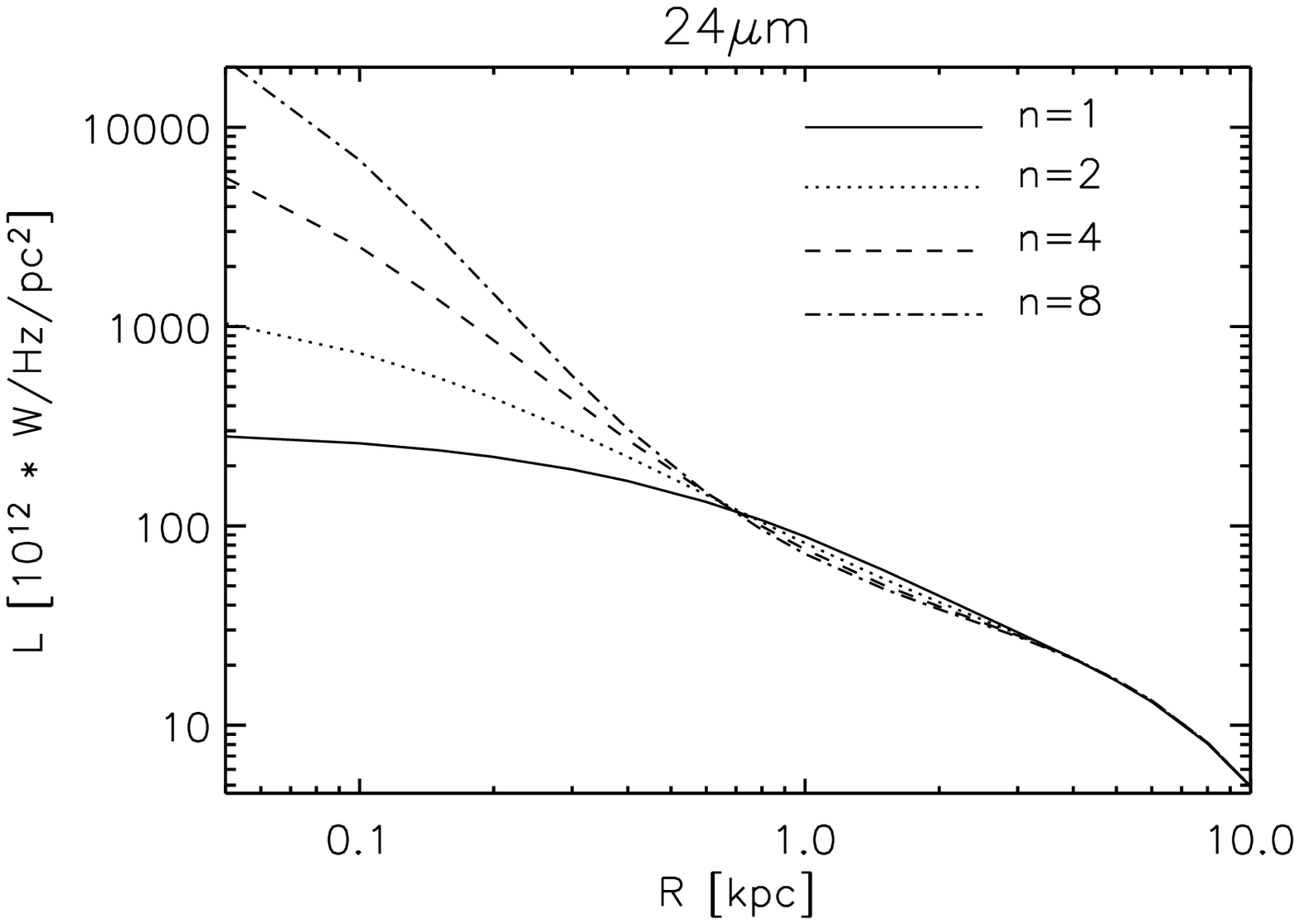}
\caption{Spatially integrated diffuse dust emission SEDs (top) and 
24\,${\mu}$m radial profiles along the major-axis (bottom) of a model galaxy
(containing a young stellar disk, an old stellar disk - with a B-band
exponential scale-length of 5.67\,kpc - and a bulge with $R^{eff}=1.3\,$kpc). 
The parameters of the model were set to $SFR=1\,{\rm M}_{\odot}/$yr, $old=1$, 
$\tau^f_B=4.0$, $B/D=0.33$. The 
different lines correspond to different S\'{e}rsic indices of the stellar 
emissivity in the bulge (see legend). The profiles are for a galaxy seen 
face-on.}
\label{fig:dust_emission}
\end{figure}

To check this effect we used the codes from Popescu et al. (2011) to perform 
calculations for dust heating and emission 
for a model galaxy having bulges described by different S\'{e}rsic indices 
($n=1,2,4,8$). We chose a typical value for the bulge-to-disk ratio, of 
$B/D=0.33$. We also considered a galaxy having the unit luminosities (as
described in our model - see Sect.~\ref{sec:library} and Popescu et al. 2011) 
for the old and young stellar populations ($SFR=1\,{\rm M}_{\odot}/$yr, $old=1$) and
$\tau^f_B=4.0$. The resulting integrated SEDs are shown in the top panel
of Fig.~\ref{fig:dust_emission}. One can see that with higher the S\'{e}rsic 
index, the larger the emission in the 24 to 60\,${\mu}$m wavelengths range
is. This is because of the very strong heating of the dust in the very central
region of the disk by optical photons emitted by the bulge. 
Although the total dust luminosity is almost unaffected by the shift in the
S\'{e}rsic index, the integrated mid-infrared/far-infrared colours are somewhat
boosted for galaxies containing bulges with high S\'{e}rsic index. This boost 
in colour is however moderate in
comparison with the boost due to an increase in the $B/D$ ratio, as shown in
Fig.~19 of Popescu et al. (2011). 

The stronger effect though appears in the
surface-brightness distribution, in particular in the central regions of 
galaxies containing bulges. In the bottom panel of
Fig.~\ref{fig:dust_emission}, one can see that the radial profiles of galaxies
containing bulges described by different S\'{e}rsic indices can vary by a
factor of more than 10 at 24\,${\mu}$m. This is remarkable, since dust
emission at 24\,${\mu}$m is commonly interpreted in terms of heating 
either by UV photons in star-formation regions or from accretion sources 
(AGN), whereas here the heating is dominated by old stars in the
bulge. Moreover, one can see that this enhancement occurs only within a few
hundred parsec of the centre. At these linear scales, for resolved galaxies 
beyond 10\,Mpc the excess emission due to the bulge heating will appear as a 
separate warm point-like source at the angular resolution of the Spitzer Space 
Telescope at 24\,${\mu}$m. Particular care must therefore be taken in
interpreting central sources seen in the mid-infrared towards spiral
galaxies with bulges, to distinguish between an obscured low-luminosity AGN, 
star-formation activity and the effect of the bulge we have identified here.

\section{Summary}
\label{sec:summary}

In this paper we provide a library of energy densities of the
diffuse radiation fields arising
from stars in the
main morphological components of spiral galaxies: disks, thin disks
and bulges. The library contains 630 two-dimensional spatial grids
(in cylindrical coordinates) of
energy densities of the RFs, spanning seven values of face-on central
optical depth $\tau^f_B$, 15
values of wavelength, and, for bulge RFs, four values of S\'{e}rsic
index $n$. Thus, we considered $\tau^f_B=0.1,0.3,0.5,1.0, 2.0,4.0,8.0$,
$n=1,2,4,8$ and wavelengths from $912\,\AA$ to $5\,{\mu}$m. The
radiation fields are sampled on an irregular grid with sampling intervals
in the radial direction ranging from 50\,pc in the centre to up to 2\,kpc in the
outer disk and in the vertical direction ranging from
50\,pc in the plane to up to 500\,pc in the outer halo.

The model radiation fields are those used to generate the large
library of  dust- and  PAH-emission SEDs presented in Popescu et
al. (2011). This library is self-consistently calculated with the
corresponding library of dust attenuations first presented in Tuffs et
al.  (2004) and then in updated form in 
Popescu et al. (2011). All calculations were made using a modified version of 
the ray-tracing radiative transfer code of Kylafis \& Bahcall (1987), and the 
dust model from 
Weingartner \& Draine (2001)and Draine \& Li (2007), incorporating a mixture 
of silicates, graphites and PAH molecules.  

We discussed the radial and vertical variation of the RFs, as well as
their variation with $\tau^f_B$, and, for bulges, with the S\'{e}rsic index.
We give analytic formulae for the radiation fields of a 
dustless stellar disk, as well as for a stellar disk with one
and with two dust disks. We use these to explain the trends in the profiles of 
the thin disk. We also show how the library can be self-consistently combined with the 
results from modelling the spatially integrated UV/optical/submm SEDs of 
galaxies from Popescu et al. (2011).

As a practical application we calculated the dust emission SEDs for
  galaxies having different bulge S\'{e}rsic indices. We find strong
  mid-infrared localised emission in the central regions of disks with high
  S\'{e}rsic index bulges, which can mimic a central star-formation region or 
  a mild AGN activity. Unlike these latter two cases, the
  dust is in this case heated by optical photons from the bulge.

\section{Acknowledgements}{We would like to thank the referee, Simone Bianchi, for his
  very insightful and constructive comments, and in particular for encouraging us
  to provide the analytic solutions}.

\appendix
\section{Derivation of an analytic solution for radiation fields 
in the mid-plane of an azimuthally symmetric infinitely thin 
dustless stellar disk} 
\label{app:opt_thin}

We take the stellar disk to be infinitely thin. The contribution to the
radiation field at a point P in the disk due to an emitting element in an annulus of the
disk at coordinates $(R^{\prime},\theta^{\prime})$ is:
                 
\begin{eqnarray}
\label{eq:dutheta}
du^{\theta} & = & \frac{\sigma\,R^{\prime}\,d{\theta}^{\prime}\,\Delta R^{\prime}}{4\,\pi\,{R^{\prime\prime}}^2\,c}
\end{eqnarray}
where
${\sigma(R^{\prime}})$ is the surface density of luminosity, $R$ is the 
galactocentric 
radius of the point P, $R^{\prime}$ is the
galactocentric radius of the annulus considered, $R^{\prime\prime}$ is the
offset of P from the emitting element in the annulus, ${\theta}^{\prime}$ is
the azimuthal offset of the emitting element with respect to P, and $c$ is the
speed of light. This geometry is illustrated for the cases that P is 
interior to the annulus in Fig.~\ref{fig:opticallythin}. It should be noted that this expression in
Eq.~\ref{eq:dutheta} applies to both these cases. The contribution to the 
radiation field at P from the entire annulus is then:
\begin{figure}
\includegraphics[scale=0.3]{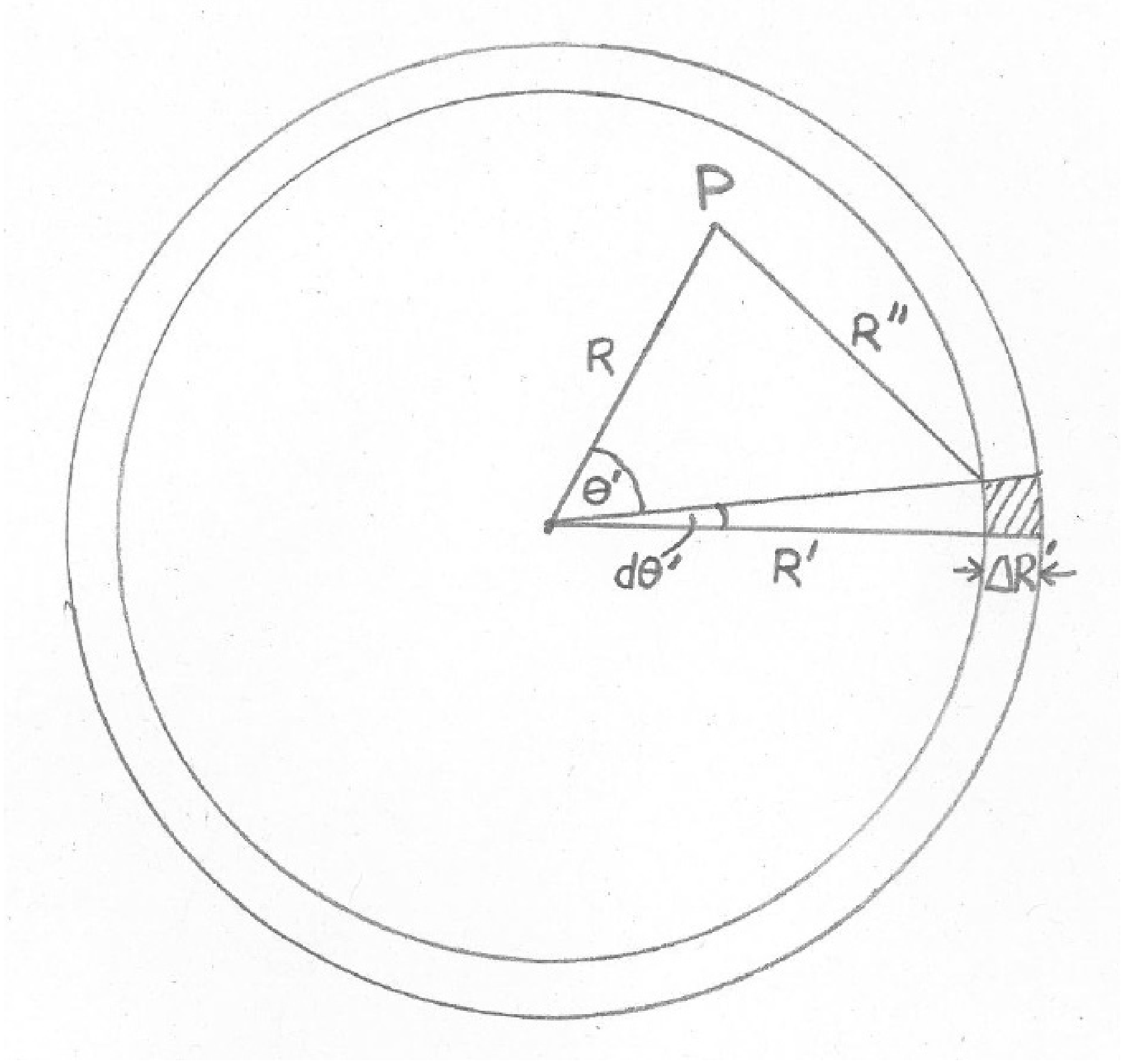}
\caption{Schematic representation of the geometry considered for the
  calculation of the radiation fields at a point P, with galactocentric radius 
  $R$, in an infinitely thin dustless stellar disk due to emitters in an 
  annulus at galactocentric radius $R^{\prime}$. The figure depicts the face-on
view point.}
\label{fig:opticallythin}
\end{figure}
 \begin{eqnarray}
\label{eq:duann}
du^{ann} & = & \int_0^{2\,\pi} du^{\theta}\,d\theta\,\, =\,\, \frac{\sigma\,R^{\prime}\,\Delta
  R^{\prime}}{4\,\pi\,c}\int_0^{2\,\pi}\frac{d\theta^{\prime}}{{R^{\prime\prime}}^2}
\end{eqnarray}
Applying the cosine rule:
\begin{eqnarray}
\label{eq:rpp}
{R^{\prime\prime}}^2 & = & R^2 + {R^{\prime}}^2 - 2\,R\,R^{\prime}\,\cos({\theta}^{\prime})
\end{eqnarray}
we obtain:
\begin{eqnarray}
\label{eq:duann1}
du^{ann} & = & \frac{\sigma\,R^{\prime}\,\Delta
  R^{\prime}}{4\,\pi\,c}\,\frac{1}{R^2+{R^{\prime}}^2}\,I
\end{eqnarray}
where
\begin{eqnarray}
\label{eq:ii}
I & = & \displaystyle \int_0^{2\,\pi}\frac{d{\theta}^{\prime}}{1-\displaystyle\frac{2\,R\,R^{\prime}\,\cos({\theta}^{\prime})}{R^2+{R^{\prime}}^2}}
\end{eqnarray}
Applying the notation:
\begin{eqnarray}
\label{eq:betabeta}
\beta & = & \frac{2\,R\,R^{\prime}}{R^2+{R^{\prime}}^2}
\end{eqnarray}
the integral in Eq.~\ref{eq:ii} becomes:
\begin{eqnarray}
\label{eq:ii1}
I & = & \int_0^{2\,\pi} \frac{d{\theta}^{\prime}}{1-\beta\,\cos({\theta}^{\prime})}
\end{eqnarray}
which has the solution
\begin{eqnarray}
\label{eq:ii2}
I & = & \frac{4}{\sqrt{1-{\beta}^2}}\,I^{\prime}
\end{eqnarray}
where
\begin{eqnarray}
\label{eq:iip}
I^{\prime}   =   2\,\tan^{-1}{\left( \frac{1+\beta}{1-\beta}\right)}^{1/2} 
   - \tan^{-1}\frac{\beta}{\sqrt{1-{\beta}^2}}\,=\,\frac{\pi}{2}
\end{eqnarray} 
Substituting Eqs.~\ref{eq:iip}, \ref{eq:ii2} and \ref{eq:betabeta} in
Eq.~\ref{eq:duann1} we obtain:
\begin{eqnarray}
\label{eq:duann2}
du^{ann} & = & \frac{\sigma\,R^{\prime}\,\Delta R^{\prime}}{2\,c\,}\,\frac{1}{|{R^{\prime}}^2-R^2|}
\end{eqnarray} 
This solution applies both for annuli exterior to P (case $R^{\prime}>R$)
and for annuli interior to P (case $R^{\prime}<R$). To obtain the total
radiation field at P we integrate over $R^{\prime}$, splitting  the
integration about the singularity at $R=R^{\prime}$: 
\begin{eqnarray}\label{eq:opt_thin}
u(R) &  = & u_1(R) + u_2(R)
\end{eqnarray}
where
\begin{eqnarray}
u_1(R) & = & \int_0^R du^{ann} dR^{\prime}\\
u_2(R) & = & \int_R^\infty du^{ann} dR^{\prime}
\end{eqnarray}
The result of this integration is:
\begin{eqnarray}
\label{eq:u1}
u_1(R) & = & \displaystyle \frac{1}{4\,{\pi}\,c\,R^2}\int_0^R
f(\frac{R'}{R})\,2\,{\pi}\,R'\,\sigma(R')\,dR'\\
\label{eq:u2}
u_2(R) & = & \displaystyle \frac{1}{4\,{\pi}\,c}\int_R^\infty
\frac{f(\frac{R}{R'})\,2\,\pi\,R'\,\sigma(R')\,dR'}{R'^2}
\end{eqnarray}
where
\begin{eqnarray}
\label{eq:fxi}
f({\xi}) & = & \displaystyle \frac{1}{1-\xi^2}
\end{eqnarray}

\section{Derivation of an analytic solution for radiation fields 
of an azimuthally symmetric dustless stellar disk with finite 
thickness}
\label{app:opt_thin_z}

It is straightforward to extend the solutions derived in
Appendix~\ref{app:opt_thin} for the radiation fields in the plane of an 
infinitely thin disk to derive expressions for radiation fields at any 
position in $(R,z)$ of a disk which has a finite thickness.

We consider now the annulus in Fig~\ref{fig:opticallythin} as lying in a plane offset $z^\prime$
from the plane of the galaxy. The contribution to the radiation field at a
point P with cylindrical coordinates $(R,z)$ due to an emitting element in the
annulus at coordinates $(R^{\prime},z^{\prime})$ is then:
\begin{eqnarray}
\label{eq:dutheta_z}
du^{\theta} & = & \frac{\eta\,R^{\prime}\,d{\theta}^{\prime}\,\Delta
  R^{\prime}\Delta z}{4\,\pi\,{R^{\prime\prime\prime}}^2\,c}
\end{eqnarray}
where $\eta$ is the volume stellar emissivity, $\Delta z$ is the thickness of
the annulus, and  $R^{\prime\prime\prime}$ is the offset (in 3D) of the point
P from the emitting element in the annulus, where
\begin{eqnarray}
\label{eq:rppp}
{R^{\prime\prime\prime}}^2 & = & {R^{\prime\prime}}^2 + {\left(z^{\prime}-z\right)}^2
\end{eqnarray}
and $R^{\prime\prime}$ is the projection of the vector joining P to the emitting
  element in the plane of the annulus. Analogous to the solution for the 
infinitely thin disk (Eq.~\ref{eq:duann}),
the contribution to the radiation field at P from the entire annulus is:
 \begin{eqnarray}
\label{eq:duann_z}
du^{ann} & = & \int_0^{2\,\pi} du^{\theta}\,d\theta\,\Delta z^{\prime}\, =\,\, \frac{\eta\,R^{\prime}\,\Delta
  R^{\prime}}{4\,\pi\,c}\int_0^{2\,\pi}\frac{d\theta^{\prime}}{{R^{\prime\prime\prime}}^2}
\end{eqnarray}
Substituting Eq.~\ref{eq:rppp} for $R^{\prime\prime}$, as given by
Eq.~\ref{eq:rpp} (applied in the plane at $z$), we obtain:
\begin{eqnarray}
\label{eq:duann1_z}
du^{ann} & = & \frac{\eta\,R^{\prime}\,\Delta
  R^{\prime}\,\Delta
  z}{4\,\pi\,c}\,\frac{1}{R^2+{R^{\prime}}^2+{\left(z^{\prime}-z\right)}^2}\,I
\end{eqnarray}
where
\begin{eqnarray}
\label{eq:ii_z}
I & = & \displaystyle
\int_0^{2\,\pi}\frac{d{\theta}^{\prime}}{1-\displaystyle\frac{2\,R\,R^{\prime}\,\cos({\theta}^{\prime})}{R^2+{R^{\prime}}^2
  +{\left(z^{\prime}-z\right)}^2}}
\end{eqnarray}
Applying the notation:
\begin{eqnarray}
\label{eq:betabeta_z}
\beta & = & \frac{2\,R\,R^{\prime}}
{R^2+{R^{\prime}}^2+{\left(z^{\prime}-z\right)}^2} 
\end{eqnarray}
we recover the same form for the integral in Eq.~\ref{eq:ii_z} as previously
obtained in Eq.~\ref{eq:ii1}, which has the solution:
\begin{eqnarray}
\label{eq:ii2_z}
I & = & \frac{2\,\pi}{\sqrt{1-{\beta}^2}}
\end{eqnarray}
Substituting Eqs.~\ref{eq:ii2_z} and \ref{eq:betabeta_z} in
Eq.~\ref{eq:duann1_z} we obtain:
\begin{eqnarray}
\label{eq:duann2}
du^{ann} & = & \frac{\eta\,R^{\prime}\,\Delta R^{\prime}\,\Delta
  z^{\prime}}{2\,c\,}\,g(R,z,R^{\prime},z^{\prime})
\end{eqnarray} 
where
\begin{eqnarray}
\label{eq:g_z}
g & = &   \frac{1}{|{\sqrt{{(R+R^{\prime})}^2+{(z^{\prime}-z)}^2}}
  \sqrt{{(R^{\prime}-R)}^2+{(z^{\prime}-z)}^2}|}
\end{eqnarray}
We note that the expression for $du^{ann}$, given by Eqs.~\ref{eq:duann2} and 
\ref{eq:g_z} reduces to Eq.~\ref{eq:duann1} for $z=z^{\prime}=0$.

Integrating over $R^{\prime}$ and $z^{\prime}$ we obtain a general solution for
the radiation field energy density arising from any distribution
of stellar emissivity with cylindrical symmetric geometry in the optically
thin limit:
\begin{eqnarray}
\label{eq:u_z}
u(R,z) & = & \displaystyle \frac{1}{2\,c}\int_{-\infty}^{\infty}\int_0^{\infty}
g(R,z,R^{\prime},z^{\prime})\,R'\,\eta(R',z^{\prime})\,dR'dz^{\prime}
\end{eqnarray}

\section{Derivation of an analytical solution for radiation fields 
in the mid-plane of an azimuthally symmetric stellar disk 
with one dust disk} 
\label{app:onedust} 
 
\begin{figure}
\includegraphics[scale=0.3]{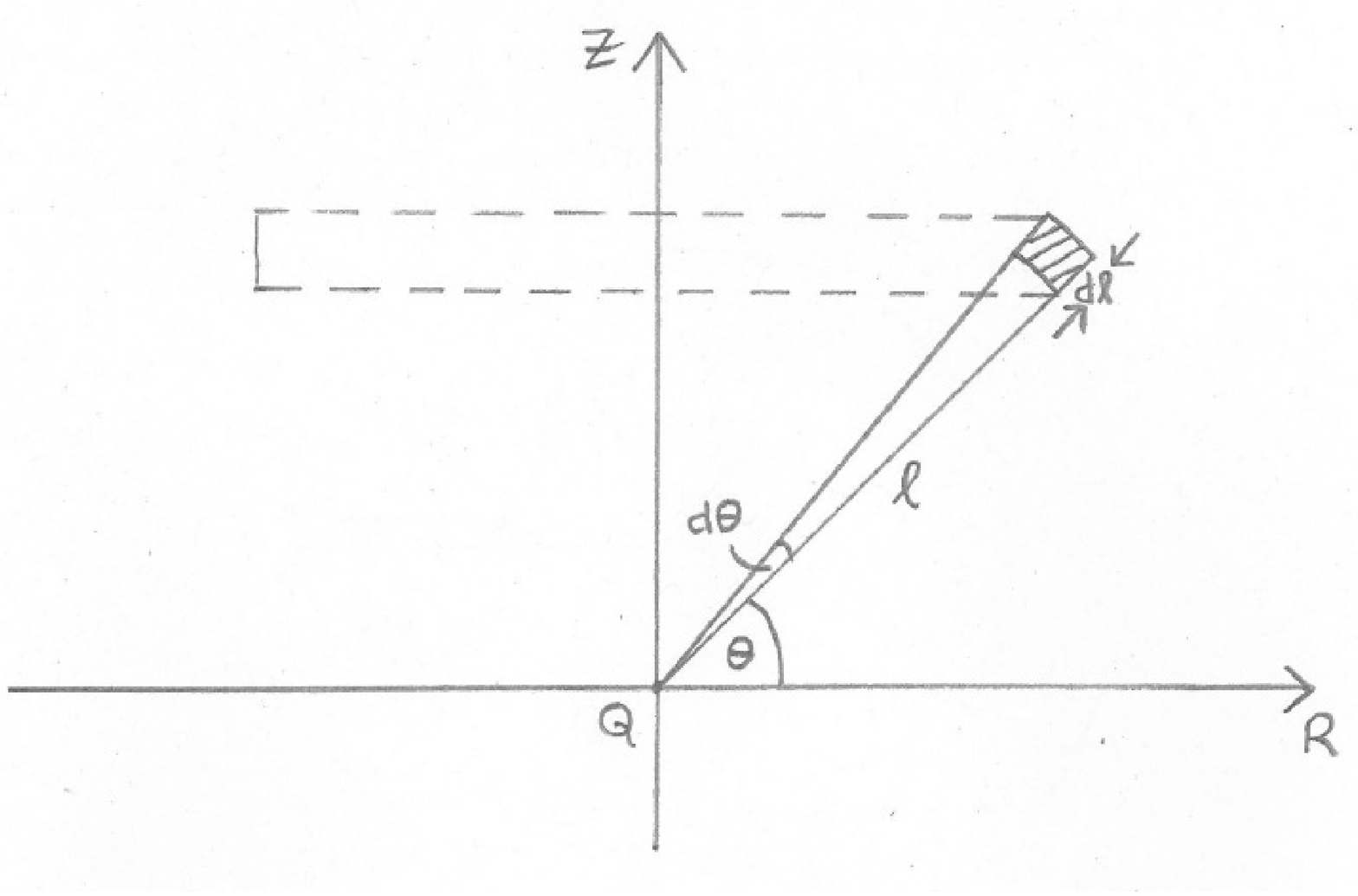}
\caption{Schematic representation of the geometry considered for the
  calculation of the radiation fields at a point Q in the mid-plane of the 
disk and at a galactocentric radius of $R$. Radiation arriving at Q
is emitted from an annulus at distance $l$, with angle $\theta$ subtended at 
the plane. The polar axis of the annulus passes through Q. The figure depicts
the edge-on view point.}
\label{fig:optically_thick}
\end{figure}

We take the stellar and dust disks to be both exponentially stratified (in
vertical direction) while having an arbitrary distribution in radial
direction. This allows us to write the volume 
stellar emissivity $\eta(R,z)$ and the extinction coefficient
$\kappa(R,z)$\footnote{Here we defined the extinction coefficient as the
  probability of absorption or scattering of a photon per unit length.} as:

\begin{eqnarray}
\label{eq:eta}
\eta(R,z) & = & \eta_{z_0}(R)\,\exp\left(-\frac{z}{z_s}\right)\\
\label{eq:kappa}
\kappa(R,z) & = & \kappa_{z_0}(R)\,\exp\left(-\frac{z}{z_d}\right)
\end{eqnarray}
where $z_s$ and $z_d$ are the scaleheights of the stellar and dust disks,
respectively, and $\eta_{z_0}(R)$ and $\kappa_{z_0}(R)$ are the mid-plane 
stellar emissivity and dust opacity. 

Our aim is to derive an analytic expression of the radiation
fields in the mid-plane of the disk in the optically thick limit. The term
optically thick is used in the sense that the terms $\eta_{z_0}(R)$ and 
$\kappa_{z_0}(R)$ are taken to be constants over the typical range in $R$ that 
photons contributing to the mid-plane energy density actually travel. This 
allows the validity of the solution to be maintained even when the disk becomes 
transparent in the vertical direction, provided that the scale in radius over
which the opacity of the dust disk changes is much larger than the scaleheight 
of the dust disk. 

The contribution to the mid-plane energy density due to direct light from stars 
in a volume element within a
thin shell of radius $(l,l+dl)$, and angle $(\theta,\theta+d\theta)$ 
(see Fig.~\ref{fig:optically_thick}) is
\begin{eqnarray}
\label{eq:du}
du & =  & u(l,\theta)\,dl\,d\theta  \\ \nonumber
   & = & \frac{\eta(l)\,dV}{4\,\pi\,l^2\,c}\,\exp\left(-\tau(l)\right)
\end{eqnarray}
where
\begin{eqnarray}
\label{eq:dV}
dV & = & 2\,\pi\,l^2\,\cos(\theta)\,d\theta\,dl\\
\label{eq:etal}
\eta(l) & = & \eta_{z_0}(R)\,\exp\left(-\frac{l\sin(\theta)}{z_s}\right)\\
\label{eq:taul}
\tau(l) & = & \kappa_{z_0}(R)\int_0^l \exp\left(\frac{-l\,\sin(\theta)}{z_d}\right)\,dl\\
\nonumber
&  = &
\frac{\kappa_{z_0}(R)\,z_d}{\sin(\theta)}\,\left[1-\exp\left(-\frac{l\sin(\theta)}{z_d}\right)\right]
\end{eqnarray}
We make the notations:
\begin{eqnarray}
\label{eq:alpha}
\alpha & = & \frac{\kappa_{z_0}(R)\,z_d}{\sin(\theta)}\\
\label{eq:beta}
\beta & = & \frac{\sin(\theta)}{z_d}\\
\label{eq:gamma}
\gamma & = & \frac{\sin(\theta)}{z_s}
\end{eqnarray}
Substituting Eqs.~\ref{eq:dV}, \ref{eq:etal} and \ref{eq:taul} into 
Eq.~\ref{eq:du}, using the notations from Eqs.~\ref{eq:alpha}, \ref{eq:beta}
and \ref{eq:gamma} and integrating over dl we obtain:
\begin{eqnarray}
\label{eq:du1}
u_{\theta}\,d\theta & = & d\theta \int_0^{\infty} u(l,\theta)\,dl\\ \nonumber
        & = & d\theta\,\frac{\eta_{z_0}(R)\,\cos(\theta)\,\exp\left(-\alpha\right)}{2\,c}\,I(\theta,\alpha,\beta,\gamma)
\end{eqnarray}
where
\begin{eqnarray} 
\label{eq:i}
I(\theta,\alpha,\beta,\gamma) & = & \int_0^{\infty}\exp \left(-\gamma\,l\right)\,
\exp\left(\alpha\,e^{-\beta\,l}\right)\,dl
\end{eqnarray}
By further manipulating Eq.~\ref{eq:i} we obtain:
\begin{eqnarray} 
\label{eq:i1}
I(\theta,\alpha,\beta,\gamma) & = & \frac{\alpha^{-x}}{\beta}\,I^{\prime}(\alpha,\beta)
\end{eqnarray}
where
\begin{eqnarray}
\label{eq:x}
x & = & \frac{z_d}{z_s}\\
\label{eq:ip}
I^{\prime} & = & \int_0^{\alpha} \xi^{x-1}\,e^{\xi}\,d\xi
\end{eqnarray}
The integral defining $I^{\prime}$ is analytic if $x$ is a non-negative
integer. The simplest solution is for $x=1$, which corresponds to the dust disk
having the same scale-height as the stellar disk. In this case  Eq.~\ref{eq:i1}
becomes:
\begin{eqnarray}
\label{eq:i2}
I(\theta,\alpha,\beta,\gamma) & = & \frac{\exp\left(\alpha\right)-1}{\kappa_{z_0}}
\end{eqnarray}
Substituting Eq.~\ref{eq:i2} into Eq.~\ref{eq:du1} and integrating over
$\theta$ 
we obtain the analytic solution for the energy density in the mid-plane due to
direct light from stars:
\begin{eqnarray}
\label{eq:u3}
u_{z_0}^{dl}(R) & = & 2 \int_0^{\pi/2} u_{\theta}\,d\theta\\ \nonumber
  & = & \frac{\eta_{z_0}(R)}{\kappa_{z_0}(R)\,c}\, \left[1-I^{\prime\prime}(R)\right]
\end{eqnarray}
where
\begin{eqnarray}
\label{eq:ipp}
I^{\prime\prime}(R) & = & \int_0^{\pi/2}\cos(\theta)\,\exp\left(-\frac{\kappa_{z_0}(R)\,z_d}{\sin(\theta)}\right)\,d\theta
\end{eqnarray}
From Eqs.~\ref{eq:u3} and \ref{eq:ipp} one sees that for very optically thick
solutions, in which the range of photons is small in $z$ as well as in $R$,
$u_{z_0}^{dl}$ tends toward:
\begin{eqnarray}
\label{eq:uthick}
u_{z_0}^{dl} & = & \frac{\eta_{z_0}(R)}{\kappa_{z_0}(R)\,c}
\end{eqnarray}
which, if the scalelength of the dust and stellar emissivity distribution is
the same, would have a flat distribution in $R$. In this optically thick limit
one can also provide an analytical expression for the combination of direct and
scattered light, since the scattered light will to a good approximation
be re-absorbed and re-scattered close to the position of the originating
stars. Under these circumstances the energy density of the radiation fields due
to first order scattered light $u^{sc,1}$ is:
\begin{eqnarray}
\label{eq:usc1}
u^{sc,1} & = & A\,u^{dl}
\end{eqnarray}
where $A$ is the albedo. 
The total energy density of the radiation fields due to both direct and
first order scattered light is:
\begin{eqnarray}
\label{eq:utotal_first}
u = u^{dl} + u^{sc,1}
\end{eqnarray}
By substituting Eq.~\ref{eq:usc1} in Eq.~\ref{eq:utotal_first} we obtain:
\begin{eqnarray}
\label{eq:utotal1_first}
u & = & u^{dl}\left(1+A\right)
\end{eqnarray}
Combining Eq.~\ref{eq:uthick} with Eq.~\ref{eq:utotal1_first} we can obtain an 
accurate solution for the  mid-plane radiation fields due to direct and first
order scattered light for the very optically thick case: 
\begin{eqnarray}
u_{z_0} & = & \frac{\eta_{z_0}(R)\,\left(1+A\right)}{\kappa_{z_0}(R)\,c}
\end{eqnarray}
Following from Eq.~\ref{eq:usc1}, the energy density due to the nth order scattered
light $u^{sc,n}$ is then given by:
\begin{eqnarray}
\label{eq:uscn}
u^{sc,n} & = & A\,u^{sc,n-1}\,\, =\,\, A^n\,u^{dl}
\end{eqnarray}
The total energy density of the radiation fields due to both direct and
scattered light is:
\begin{eqnarray}
\label{eq:utotal}
u = u^{dl} + \sum_{n=1}^{\infty} u^{sc,n}
\end{eqnarray}
By substituting Eq.~\ref{eq:uscn} in Eq.~\ref{eq:utotal} we obtain:
\begin{eqnarray}
\label{eq:utotal1}
u & = & u^{dl}\sum_{n=0}^{\infty}A^n\,\, =\,\, \frac{u^{dl}}{1-A}
\end{eqnarray}
Combining Eq.~\ref{eq:uthick} with Eq.~\ref{eq:utotal1} we can obtain an 
accurate solution for the total 
mid-plane radiation fields for the very optically thick case, in which the 
range of photons is small compared to $z_d$ as well as being small compared 
to the scale over which opacity varies with $R$: 
\begin{eqnarray}
\label{eq:main}
u_{z_0} & = & \frac{\eta_{z_0}(R)}{\left(1-A\right)\,\kappa_{z_0}(R)\,c}
\end{eqnarray}
Finally we can write an approximate solution for the total mid-plane energy
density for a disk with a scale-height equal to the scale-height of the
stars, in which we relaxed the condition that the disk is optically thick in
$z$, but in which we retained the condition that the disk is still optically
thick in $R$. For this we combine Eq.~\ref{eq:utotal1_first} (for first order
scattering) or Eq.\ref{eq:utotal1} (for total scattering) with the exact
solution for direct light from Eq.~\ref{eq:u3}  to obtain:
\begin{eqnarray}
\label{eq:final_firstorder}
u_{z_0}(R) & = &
\frac{\eta_{z_0}(R)\left(1+A\right)}{\kappa_{z_0}(R)\,c}\, \left[1-I^{\prime\prime}(R)\right]
\end{eqnarray} 
for direct plus first order scattering or 
\begin{eqnarray}
\label{eq:final_allorders}
u_{z_0}(R) & = &
\frac{\eta_{z_0}(R)}{\left(1-A\right)\,\kappa_{z_0}(R)\,c}\, \left[1-I^{\prime\prime}(R)\right]
\end{eqnarray} 
for direct plus total scattered light.

Because in most real cases the solution is never completely optically thick, we
include in the formula a further factor, $f_{esc}$ ($f^1_{esc}$ for first order
scattering), which corrects for the
extra escaping radiation which is not accounted by the approximation made for
the scattering:

\begin{eqnarray}
\label{eq:final_firstorder_corr}
u_{z_0}(R) & = &
f^1_{esc}\,\frac{\eta_{z_0}(R)\left(1+A\right)}{\kappa_{z_0}(R)\,c}\, \left[1-I^{\prime\prime}(R)\right]
\end{eqnarray} 
for direct plus first order scattering, or 
\begin{eqnarray}
\label{eq:final_allorders_corr}
u_{z_0}(R) & = &
f_{esc}\,\frac{\eta_{z_0}(R)}{\left(1-A\right)\,\kappa_{z_0}(R)\,c}\,
\left[1-I^{\prime\prime}(R)\right]
\end{eqnarray} 
for direct plus total scattered light. The factors $f_{esc}$ and $f^1_{esc}$ 
depend
on $\tau^f_B$ and ${\lambda}$. In the very optically thick case $f_{esc}$ and
$f^1_{esc}$ both take the value unity. In all other cases they can only be 
derived empirically, by
calibrating them on radiative transfer calculations. In our formulation we
  consider $f_{esc}$ (or $f^1_{esc}$) a single number (independent of
  galactocentric position). This provides an estimate of the goodness of 
  our approximation (for scattered light) for the centre of the disk,
  where the higher opacity is encountered, and therefore the best agreement is
  expected. Thus, the calibration is done by scaling the profiles at $R=0$. 
The tables with the values of the $f_{esc}$ and $f^1_{esc}$ for $\tau^f_B=3.5$
are given in Table~\ref{tab:onedisk}.

\begin{table}
\caption{The values of $f_{esc}$ and $f^1_{esc}$ for the case of one dust disk
  with $\tau^f_B=3.5$}
\label{tab:onedisk}
\begin{tabular}{l|ll}
\hline
$\lambda$ & $f^1_{esc}$ & $f_{esc}$\\
\hline
912 & 0.980 & 0.976 \\
1350 & 0.976 & 0.956\\
1500 & 0.976 & 0.926\\
1650 & 0.976 & 0.926\\
2000 & 0.976 & 0.833\\
2200 & 0.980 & 0.833\\
2500 & 0.976 & 0.714\\
2800 & 0.966 & 0.658\\
3600 & 0.952 & 0.595\\
\hline
\end{tabular}
\end{table}

\section{Derivation of an analytical solution for radiation fields 
in the mid-plane of an azimuthally symmetric stellar disk 
with two dust disks}
\label{app:twodust} 

We take the stellar disk to have the same form as in
Eq.~\ref{eq:eta}.  The two dust disks have also an exponential distribution in
the vertical direction, but an arbitrary distribution in radial direction. 
This allows us to write the extinction coefficient $\kappa(R,z)$ as a sum of the
extinction coefficients of each dust disk:
\begin{eqnarray}
\label{eq:kappa2}
\kappa(R,z) & = & \kappa_1(R,z) + \kappa_2(R,z) =\\ \nonumber
            & = & \kappa_{z_{01}}(R)\,\exp\left(-\frac{z}{z_{d1}}\right) +
            \kappa_{z_{02}}(R)\,\exp\left(-\frac{z}{z_{d2}}\right)\\
\kappa_{z_{01}}(R) & = & \frac{\tau_1}{2\,z_{d1}}\\
\kappa_{z_{02}}(R) & = & \frac{\tau_2}{2\,z_{d2}}\\
\tau & = & \tau_1+\tau_2
\end{eqnarray}
where $\kappa_1(R,z)$ and $\kappa_2(R,z)$ are the extinction coefficients of
the first and second dust disks, respectively, $\kappa_{z_{01}}(R)$ and
$\kappa_{z_{02}}(R)$ are the mid-plane dust extinction coefficents of the first
and second dust disks, respectively, and $z_{d1}$ and $z_{d2}$ are the
scaleheights of the dust disks.

Following the same calculations as in Sect.~\ref{app:onedust}, we obtain
the same equations as \ref{eq:du}, \ref{eq:dV}, \ref{eq:etal}, but a different
form from \ref{eq:taul}, namely:
\begin{eqnarray}
\label{eq:taul2}
\tau(l) & = & 
\kappa_{z_{01}}(R)\int_0^l\exp\left(\frac{-l\,\sin(\theta)}{z_{d1}}\right)\,dl\\
\nonumber
 &+ & 
\kappa_{z_{02}}(R)\int_0^l\exp\left(\frac{-l\,\sin(\theta)}{z_{d2}}\right)\,dl\\
\nonumber
&  = &
\frac{\kappa_{z_{01}}(R)\,z_{d1}}{\sin(\theta)}\,\left[1-\exp\left(-\frac{l\sin(\theta)}{z_{d1}}\right)\right]\\
\nonumber
 &+ & 
\frac{\kappa_{z_{02}}(R)\,z_{d2}}{\sin(\theta)}\,\left[1-\exp\left(-\frac{l\sin(\theta)}{z_{d2}}\right)\right]
\end{eqnarray}
We make the notations:
\begin{eqnarray}
\label{eq:alpha1}
\alpha_1 & = & \frac{\kappa_{z_{01}}(R)\,z_{d1}}{\sin(\theta)}\,\, =\,\, \frac{\tau_1}{2\,\sin\left(\theta\right)}\\
\label{eq:alpha2}
\alpha_2 & = & \frac{\kappa_{z_{02}}(R)\,z_{d2}}{\sin(\theta)}\,\, =\,\, \frac{\tau_2}{2\,\sin\left(\theta\right)}\\
\label{eq:beta1}
\beta_1 & = & \frac{\sin(\theta)}{z_{d1}}\\
\label{eq:beta2}
\beta_2 & = & \frac{\sin(\theta)}{z_{d2}}\\
\label{eq:gamma1}
\gamma & = & \frac{\sin(\theta)}{z_s}\\
\label{eq:alpha_total}
\alpha & = & \alpha_1 + \alpha_2 
\end{eqnarray}
With the new notation, the analogous of Eq.~\ref{eq:du1} becomes:
\begin{eqnarray}
\label{eq:du1_total}
u_{\theta}\,d\theta & = & d\theta \int_0^{\infty} u(l,\theta)\,dl\\ \nonumber
        & = & d\theta\,\frac{\eta_{z_0}(R)\,\cos(\theta)\,\exp\left(-\alpha\right)}{2\,c}\,I(\theta,\alpha_1,\alpha_2,\beta_1,\beta_2,\gamma)
\end{eqnarray}
where
\begin{eqnarray} 
\label{eq:i_total}
I & = &
\int_0^{\infty}\exp \left(-\gamma\,l\right)\,
\exp\left(\alpha_1\,e^{-\beta_1\,l}\right)\,\exp\left(\alpha_2\,e^{-\beta_2\,l}\right)dl
\end{eqnarray}
By further manipulating Eq.~\ref{eq:i_total} we obtain:
\begin{eqnarray} 
\label{eq:i1_total}
I & = & \int_0^1\xi^{\,\gamma-1}\,
\exp\left(\alpha_1\,\xi^{\,\beta_1}+\alpha_2\,\xi^{\,\beta_2}\right)\,d\xi
\end{eqnarray}
For
\begin{eqnarray}
x & = & \frac{z_{d2}}{z_{d1}} 
\end{eqnarray}
\begin{eqnarray}
\label{eq:i2_total}
I & = &
\left(\frac{x}{1+x}\right)\,\left(\frac{\omega_1}{2\,\kappa_{z_{01}}}\right)\,I^{\prime\prime}
\end{eqnarray}
where
\begin{eqnarray}
\label{eq:ipp_total}
I^{\prime\prime} & = & \int_0^{\frac{\tau}{2\sin\left(\theta\right)}} \xi^{\,-\frac{1}{1+x}}\,\exp\left(\omega_1\,\xi^{\frac{x}{1+x}}+\omega_2\,\xi^{\frac{1}{1+x}}\right)\,d\xi\\
\omega_1(\theta) & = &
\frac{\tau_1}{\tau^{\frac{x}{1+x}}\,\left(2\,sin\left(\theta\right)\right)^{\frac{1}{1+x}}}\\
\omega_2(\theta) & = &
\frac{\tau_2}{\tau^{\frac{1}{1+x}}\,\left(2\,sin\left(\theta\right)\right)^{\frac{x}{1+x}}}
\end{eqnarray}
By substituting Eqs.~\ref{eq:ipp_total} and \ref{eq:i2_total} into Eq.~\ref{eq:du1_total} and
integrating over $\theta$ we obtain the analytic solution for the energy
density in the mid-plane due to direct light from stars:
\begin{eqnarray}
\label{eq:utotal_dl}
u_{z_0}^{dl}(R) & = & 2 \int_0^{\pi/2} u_{\theta}\,d\theta\\ \nonumber
              & = & \left(\frac{x}{1+x}\right)\frac{\eta_{z_0}}{\kappa_{z_{01}}\,c}\int_0^{\pi/2}d\theta\cos\left(\theta\right)\exp\left(\frac{-\tau}{2\,sin\left(\theta\right)}\right)\,\omega_1(\theta)\,I^{\prime\prime}
\end{eqnarray}
The formula in Eq.~\ref{eq:utotal_dl} can be reduced to Eq.~\ref{eq:uthick}
describing the case of a stellar disk with a single dust disk if one considers:
$\tau_2=0$, $\omega_2=0$, $\tau_1=\tau$, $\omega_1=\omega$ and
$\kappa_{z_{01}}=\kappa_{z_0}$.  

Similarly to the case of one single dust disk, the solution including scattered
light can be obtained by altering the solution for direct light with a term
including the albedo $A$. Thus, the solution for the total mid-plane energy
density is:
\begin{eqnarray}
\label{eq:twodust_final_firstorder_corr}
u_{z_0}(R) & = &
f^1_{esc}\,\left(1+A\right)\left(\frac{x}{1+x}\right)\frac{\eta_{z_0}}{\kappa_{z_{01}}\,c}\\\nonumber
& \times & \int_0^{\pi/2}d\theta\cos\left(\theta\right)\exp\left(\frac{-\tau}{2\,sin\left(\theta\right)}\right)\,\omega_1(\theta)\,I^{\prime\prime}
\end{eqnarray}
for the direct plus first order scattering or
\begin{eqnarray}
\label{eq:twodust_final_allorders_corr}
u_{z_0}(R) & = &
f_{esc}\,\left(\frac{1}{1-A}\right)\left(\frac{x}{1+x}\right)\frac{\eta_{z_0}}{\kappa_{z_{01}}\,c}\\\nonumber
& \times & \int_0^{\pi/2}d\theta\cos\left(\theta\right)\exp\left(\frac{-\tau}{2\,sin\left(\theta\right)}\right)\,\omega_1(\theta)\,I^{\prime\prime}
\end{eqnarray}
for the direct plus total scattered light.

In the case of a highly optically thick disk, this reduces to an exact 
expression:
\begin{eqnarray}
\label{eq:twodust_final_allorders_corr_optthick}
u_{z_0}(R) & = &
\left(\frac{1}{1-A}\right)\left(\frac{x}{1+x}\right)\frac{\eta_{z_0}}{\kappa_{z_{01}}\,c}
\end{eqnarray}
\begin{table}
\caption{The values of $f_{esc}$ for the case of two dust disks with 
$\tau^f_B=3.5$}
\begin{tabular}{l|ll}
\hline
$\lambda$ & $f_{esc}$ \\
\hline
912 & 0.970 \\
1350 & 0.926\\
1500 & 0.909\\
1650 & 0.862\\
2000 & 0.826\\
2200 & 0.870\\
2500 & 0.699\\
2800 & 0.645\\
3600 & 0.581\\
\hline
\end{tabular}
\end{table}

\end{document}